\documentclass[a4paper,11pt]{article}

\usepackage[latin1]{inputenc}
\usepackage{authblk}
\usepackage{amsmath}
\usepackage{amsfonts}
\usepackage{amssymb}
\usepackage{graphicx}
\usepackage[toc, page]{appendix}
\usepackage{dcolumn}
\usepackage{caption}
\usepackage{float}
\usepackage{bm}
\usepackage{multicol}
\usepackage{multirow}
\usepackage{enumerate}
\usepackage[usenames]{color}
\usepackage[colorlinks=true]{hyperref}
\usepackage{pdfpages}

\usepackage[usenames]{color}
\usepackage[colorlinks=true]{hyperref}
\usepackage{pdfpages}

\title{Statistical Isotropy violation of CMB polarization sky due to Lorentz boost}
\author[1]{Suvodip Mukherjee \thanks{suvodip@iucaa.ernet.in}}
\author[1,2]{Aritra De\thanks{de.aritra1@gmail.com}}
\author[1]{Tarun Souradeep \thanks{tarun@iucaa.ernet.in}}
\affil[1]{IUCAA, Post Bag 4, Ganeshkhind, Pune-411007, India}
\affil[2]{National Institute of Science Education and Research\\
 Institute of Physics Campus, Sachivalaya Marg, PO: Sainik School, Bhubaneswar - 751 005, India}

\begin{document}

\maketitle{} 

\pagenumbering{arabic}
\thispagestyle{plain}
\markright{}

\begin{abstract}
In the frame of a moving observer, the Cosmic Microwave Background (CMB) fluctuation exhibits violation of Statistical Isotropy (SI). The SI violation effect from our local motion on CMB temperature fluctuation has been measured in the recent Planck results \cite{Planck4}. We calculate the effect of our local motion with velocity $(\beta \equiv |\boldsymbol v|/c= 1.23 \times 10^{-3})$ on CMB polarization field. The Lorentz transformation of the polarization field leads to  aberration  in the direction of incoming photons and also modulation of the Stokes parameters, which results in mixing of power between different  CMB multipoles. We show that for small values of $\beta$, the effect on the angular power spectra that corresponds to the diagonal terms in the spherical harmonic space is at $O(\beta^2)$. But non-zero off-diagonal terms at the linear order in $\beta$ could provide a measurable signature of SI violation in the Bipolar Spherical Harmonic (BipoSH) representation. We also calculate the measurability of $\beta$ from polarization maps from experiments like Planck and PRISM. It is possible to measure $\beta$ from the ideal, cosmic variance limited  BipoSH spectra of $EE$, $TE$, $BB$, but not in $EB$ and $TB$. With the instrumental noise of and angular resolution of Planck, it is not possible to measure $\beta$ with high statistical significance  from BipoSH spectra of polarization. PRISM can measure $\beta$ with high significance in both $EE$ and $TE$ BipoSH spectra, but not in $BB$, $EB$ and $TB$ BipoSH spectra. 
\end{abstract}

\section{Introduction}
Cosmic Microwave Background (CMB) is a very important probe of our universe. It has opened an era of precision cosmology with a  number of important experiments like COBE, WMAP, Planck, BOOMERanG, ACT, SPT and many others. Recent measurements from Planck of the CMB temperature power spectrum matches well with the minimal $\Lambda$CDM model \cite{Planck}. Another observable is the pattern of linear polarization on CMB sky. Study of CMB polarization maps will provide more information about lensing, inflation models and  gravitational waves.   \\
The measurement of the CMB dipole implies that we observe the CMB from the reference frame moving at $v= 369.0 \pm 0.9 \,\mathrm {km\, s^{-1}}$ in the direction, $(l,b)= (263.99^o, 48.26^o)$ in Galactic coordinates \cite{WMAP}. This is an unavoidable systematic effect for any CMB observations. The CMB measurements have achieved the sensitivity and resolution where the subtle effect of this motion on the Statistical Isotropy (SI) of correlations of fluctuations at multipoles beyond the dipole are also measurable. Recently, Planck \cite{Planck1} has also measured the SI violation induced by the local motion. CMB polarization map is also an important probe for measuring the presence of any Statistical non-Isotropy (nSI) features of CMB sky. A measure of  SI violation are the non zero coefficients of Bipolar Spherical Harmonics (BipoSH) representation introduced in CMB temperature measurements by Hajian \& Souradeep \cite{ts} and extended to  CMB polarization by Basak, Hajian \& Souradeep \cite{tsb}.\\
\\
 We study the effect of our local motion on the CMB polarization field. In this work we estimate the signature of SI violation in the CMB polarization field due to the Lorentz transformation corresponding to our moving reference frame (with $\beta \equiv |\boldsymbol {v}|/c = 1.23 \times 10^{-3}$). Applied to the two point correlations function of polarization field, this transform leads to  a mild corrections at $O(\beta^2)$ to the angular power spectra $C_l^{EE}$, $C_l^{TE}$, $C_l^{BB}$ and $C_l^{TT}$ values. However, it  induces mixing of power at different CMB multipoles at linear in $\beta$ and hence has a much stronger signature in SI violation. The effect of Doppler boost on the  covariance matrix of temperature and polarization of the CMB have been studied earlier by Challinor \& Leeuwen \cite{challinor} where they compute the effect on diagonal and off-diagonal terms of the covariance matrix specifically for a white spectra of $C_l$. Detectability of our local motion with Planck like instrumental noise were also computed by Kosowsky \& Kahniashvili \cite{kosowsky} for temperature and by Amendola et al. \cite{amendola} for temperature and polarization.  We present a comprehensive study of the effect of Doppler boost on CMB polarization in the  BipoSH representation of SI violation for the best fit $\Lambda CDM$ model. We also derive  the statistics of BipoSH coefficients for polarization and assess the measurability of Doppler boost magnitude and direction from Planck and PRISM by using minimum variance estimator \cite{hu2, duncan, duncan2} in the BipoSH representation. \\
 The paper is organised as follows. In Section 2, we briefly review some relevant background concepts and results used for obtaining our results. In Section 3, we  calculate the effect of local motion on angular power spectra and on BipoSH coefficients. Statistics of BipoSH coefficients for polarization are derived in Section 4. In Section 5, we estimate the detectability of $\beta$ from the BipoSH coefficients of temperature and polarization for Planck and PRISM using minimum variance estimator. Discussion and conclusion of our work are provided in Section 6.
\section{Brief review of relevant concepts and basic results}
In this section we briefly review some relevant concepts and basic results, which we need to use for this work. We discuss the basic result of Lorentz transformation on polarization vector, formalism of CMB polarization and BipoSH representation.
\subsection{Lorentz transformation of Stokes parameter}
  Polarization of the CMB photons at a direction $\hat{n}$ are expressed in terms of Stokes parameters $(I,Q,U,V)$, with respect to the set of basis vectors $(\hat e_1, \hat e_2)$ defined on the local sky patch in the direction, $\hat n$.  The polarization of the electromagnetic wave is defined by the direction of the electric field. We can define the electric field vector of the wave as $\vec {\mathcal{E}}(\hat{n})= \mathcal{E}_1 \hat{e_1} + \mathcal{E}_2 \hat{e_2}$, where $(\hat e_1, \hat e_2, \hat n)$ forms a right handed orthonormal system. The Stokes parameters can be defined as,
\begin{align} \label{eq1}
\begin{split}
I&= {|\mathcal{E}_1|^{2} + |\mathcal{E}_2|^{2}},\\
Q&= {|\mathcal{E}_1|^{2} - |\mathcal{E}_2|^{2}},\\
U&={ 2\rm Re(\mathcal{E}_1^{*}\mathcal{E}_2)},\\
V&= {2\rm Im(\mathcal{E}_1^{*}\mathcal{E}_2)}.
\end{split}
\end{align}
CMB polarization  is induced by the Thomson scattering of CMB photons with the electrons, which generates only linear polarization, i.e. $V=0$. Hence, we consider only $I,Q,U$ in the further discussions. These Stokes parameters can also be expressed in units of mean intensity ($\bar{\mathcal{J_\nu}}=\delta \mathcal{J_\nu}/(\mathcal{J}_\nu)_{0}$) as,
\begin{equation}\label{eq1a}
\mathcal{J}_{ij} =
 \begin{pmatrix}
  I+Q & U\\
  U & I-Q 
 \end{pmatrix},\\
\end{equation}
where, frequency dependent  intensity of the CMB field is related to CMB temperature  by Planckian distribution. 
\begin{align}\label{eq1c}
\mathcal{J}_\nu(\hat{n})= \frac{2h\nu^3}{c^2}\frac{1}{(\exp[h\nu/kT(\hat n)] -1)}\,\,.
\end{align}
The relation between the  intensity fluctuation, $\delta  \mathcal{J}_\nu$ and temperature fluctuation, $\delta T$ is
\begin{align}
 \mathcal{J}_\nu(\hat{n})- (\mathcal{J}_\nu)_0(\hat{n}) \equiv \delta \mathcal{J}_\nu(\hat{n})= \frac{d\mathcal{J}_\nu}{dT}\bigg|_{T_0} \delta T(\hat n) + \frac{1}{2}\frac{d^2\mathcal{J}_\nu}{dT^2}\bigg|_{T_0}(\delta T(\hat n))^2.
\end{align}
Performing a rotation by an angle $\phi$ on this orthonormal basis, $(\hat e_1, \hat e_2, \hat n)$, around $\hat n$, leads to change in $(\hat e_1, \hat e_2)$ as,
\begin{align} \label{eq2}
\begin{split}
\hat{e}_1^R&= \cos\phi \,\hat{e}_1 +\sin\phi \,\hat{e}_2,\\
\hat{e}_2^R&= -\sin\phi \,\hat{e}_1 +\cos\phi \,\hat{e}_2,
\end{split}
\end{align}
where we represent  the expressions in the rotated framed by a superscript $R$. This implies that the $I,Q$ and $U$ defined in eq.\eqref{eq1}, also transform as \cite{marc,zal},
\begin{align} \label{eq3}
\begin{split}
I^R=&\, I, \\
Q^R=& \, Q \,\cos2\phi + U \,\sin2\phi,\\
U^R=& \, -Q \,\sin2\phi + U\,\cos2\phi .
\end{split}
\end{align}
Hence $(Q \pm iU)$ transforms as a spin two variable,
\begin{align} \label{eq4}
(Q \pm iU)^R = e^{\mp i2\phi}(Q \pm iU).
\end{align}
Stokes parameters measured by a moving observer is related to the stationary observer through Lorentz transformation.  There are two different effects on the CMB polarization field due to the motion of the observer, \\
\\
\begin{enumerate} 
\item 
 Aberration in the direction $\hat n$, of incoming photons lead to remapping of the Stokes parameters on the sky.\\
\item
Modulation of Stokes parameters.\\
\end{enumerate}

To obtain the Lorentz transformation of CMB polarization field, 
let ${\mathbf S}^\prime$ be the CMB rest frame. 
The observer reference frame $\mathbf S$ is moving with a velocity $\boldsymbol{\beta}= \beta_0 \hat{\beta}$, with $\beta_0 = 1.23 \times 10^{-3} $ as measured by CMB dipole \cite{WMAP}.  Lorentz transformation of CMB photons lead to aberration in the direction of incoming photon.  The aberration in the direction $\hat n$, results in a remapping of $\hat n \rightarrow \hat n'$ by the relation \cite{Planck1, challinor},
\begin{align} \label{eq9}
\hat n = \frac{\hat n'. \hat \beta + \beta}{1+\hat n'. \hat \beta}\hat \beta + \frac{\hat n' - \hat n'.  \boldsymbol{\beta} \hat \beta}{\gamma (1+ \hat n'. \hat \beta)}\,.
\end{align}
The Lorentz transformation of CMB temperature field, after retaining only first order terms in $\beta$ and choosing the coordinate system in which $\hat \beta$ is along $\hat z$ axis, can be written as  \cite{Planck1}, 
\begin{align} \label{eq11}
\begin{split}
\delta T(\hat n)= \delta T'\bigg(\hat n - {\bigtriangledown}(\hat n. \hat \beta)\bigg)\bigg(1+\beta\cos\theta\bigg),
\end{split}
\end{align}
and the relation between  temperature fluctuations (excluding the dipole term) and  intensity fluctuations given by,
\begin{align} \label{eq11a}
\begin{split}
\delta T_I (\hat n) \equiv \frac{\delta \bar{\mathcal{J}_\nu}(\hat{n})}{d \bar{\mathcal{J}_\nu}/dT\bigg|_{T_0}}=  \delta T' \bigg(\hat {n} - {\bigtriangledown}(\hat {n}. \hat \beta)\bigg)\bigg(1+b_{\nu} \beta\cos\theta\bigg),
\end{split}
\end{align}
where, $\delta T_I$ is the temperature fluctuation obtained from CMB intensity fluctuation, $\theta$ is the angle between $\hat n$ and $\hat \beta$ and $b_{\nu}$ is the frequency dependent effect on Doppler boost given by,
\begin{align} \label{eq11b}
\begin{split}
b_{\nu}= \frac{\nu}{\nu_0}\mathrm{coth}\bigg(\frac{\nu}{2\nu_0}\bigg) -1,
\end{split}
\end{align}
with $\nu_0 = 57$ GHz. 
Eq.\eqref{eq11a} relates fluctuations in the CMB intensity and temperature. 
This provides the expression of Stokes parameters in terms of temperature fluctuation. So, under Lorentz transformation $Q,U$ transforms as,
\begin{align} \label{eq11d}
\begin{split}
Q(\hat{n})= Q'\bigg(\hat {n} - {\bigtriangledown}(\hat {n}. \hat \beta)\bigg)\bigg(1+b_{\nu} \beta\cos\theta\bigg),\\
U(\hat{n})= U'\bigg(\hat {n} - \bigtriangledown(\hat {n}. \hat \beta)\bigg)\bigg(1+b_{\nu} \beta\cos\theta\bigg).
\end{split}
\end{align}
These are the Lorentz transformation of Stokes parameters with a  frequency dependence given by $b_\nu$. In this paper, we choose the frequency channel $\nu = 217$ GHz  for which $b_\nu \approx 3$. This is one of the best Planck frequency channel for this search due to high resolution and low noise for the estimation of $\beta$. 
These equations relate CMB observables in the CMB rest frame (${\mathbf S}^\prime$) with moving observer frame (${\mathbf S}$) when the direction of motion is along the $\hat z$ axis and $\theta$ and $\phi$ are the usual basis in a spherical coordinate system. Statistical Isotropy (SI) in  ${\mathbf S}^\prime$  still leads to observable Statistical non-Isotropy (nSI) in ${\mathbf S}$. 

\subsection{Measures of the CMB polarization field}
The CMB temperature anisotropy sky map $\delta T (\hat n)$ can be expanded in the orthonormal space of Spherical Harmonic (SH) functions,
\begin{align}\label{eqbi1}
\delta T(\hat n)= \sum_{lm}  a_{lm}Y_{lm}(\hat n).
\end{align}
Similarly, the Stokes parameters mentioned in eq.\eqref{eq1} for the CMB polarization field, can also be expanded in terms of Spin Weighted Spherical Harmonic (SWSH) basis $_{\pm 2}Y_{lm}$, due to its transformation property mentioned in eq.\eqref{eq4}. Defining $Q \pm iU$ as \cite{marc,zal,wayne},
\begin{align} \label{eq12}
_{\pm}X(\hat n)= Q (\hat n) \pm iU (\hat n).
\end{align}
Expressing  ${}_{\pm}X(\hat n)$, in the SWSH basis \cite{marc, zal},
\begin{align} \label{eq13}
_{\pm}X(\hat n) = \sum_{l=2}^{\infty} \sum_{m=-l}^{+l}{}_{\pm}X_{lm} \,{}_{\pm 2}Y_{lm}(\hat n),
\end{align}
and in parity eigenstates we can write,
\begin{align}\label{eq14}
_{\pm}X(\hat n)= \sum_{l=2}^{\infty} \sum_{m=-l}^{+l} (E_{lm}\pm iB_{lm})  _{\pm2}Y_{lm}(\hat n),
\end{align}
where,
\begin{align}\label{eq15}
_\pm X_{lm}= E_{lm} \pm iB_{lm}.
\end{align}
Under parity transformation, $E$, $B$ and $_sY_{lm}$ transform as,
\begin{align}\label{eq16}
\begin{split}
_sY_{lm} \rightarrow (-1)^{l} _{-s}Y_{lm},\\
E_{lm}\rightarrow (-1)^{l}E_{lm},\\
B_{lm}\rightarrow (-1)^{l+1}B_{lm}.
\end{split}
\end{align}
For SI field, the angular power spectra of CMB polarization are  related to the two point correlation of $E$ and $B$, and also the cross correlation between $E$ and $T$.\begin{align}\label{eq17}
\bigg\langle Y^*_{lm}Y'_{l'm'}\bigg\rangle = C_l^{YY'}\delta_{ll'}\delta_{mm'},
\end{align}
where $Y$ and $Y'$ can be taken to be $E,B$ and $T$. Under the assumption of SI, the covariance matrix formed by $\langle Y^*_{lm}Y'_{l'm'}\rangle$, is diagonal and  non zero only for $EE$, $BB$, and $TE$. When all processes respect parity invariance, the diagonal elements correspond to the angular power spectrum, defined as, 
\begin{align}\label{eq18}
\begin{split}
\bigg\langle a^*_{lm}a_{l'm'}\bigg\rangle = C_l^{TT}\delta_{ll'}\delta_{mm'},\\
\bigg\langle E^*_{lm}E_{l'm'}\bigg\rangle = C_l^{EE}\delta_{ll'}\delta_{mm'},\\
\bigg\langle B^*_{lm}B_{l'm'}\bigg\rangle = C_l^{BB}\delta_{ll'}\delta_{mm'},\\
\bigg\langle E^*_{lm}a_{l'm'}\bigg\rangle = C_l^{TE}\delta_{ll'}\delta_{mm'}.
\end{split}
\end{align} 
 
\subsection{BipoSH representation for temperature and polarization}\label{sc1}
 In this section, we discuss the nSI part of the covariance matrix for polarization and temperature in the Bipolar Spherical Harmonics (BipoSH) representation\cite{ts, tsb}. Under the assumption that the temperature fluctuations are Gaussian random field with zero mean, we can express the statistics of this field by the two point correlation function. In the full generality the two point correlation of SH coefficients of the CMB anisotropy $\langle a_{l_1m_1} a^*_{l_2m_2}\rangle$ in SH space can be expanded in the tensor product basis of two SH spaces as, 
 \begin{align}\label{eqbi2}
\bigg\langle a_{l_1m_1} a^*_{l_2m_2}\bigg\rangle = \sum_{LM} A^{LM}_{l_1l_2}(-1)^{m_2}C^{LM}_{l_1m_1l_2-m_2} ,
\end{align}
where, $A^{LM}_{l_1l_2}$ are called the Bipolar Spherical Harmonics (BipoSH) coefficients \cite{ts} and $C^{LM}_{l_1m_1l_2m_2}$ are the Clebsch-Gordan coefficients. For nSI case, non-zero BipoSH coefficients are a complete representation of SI violation.\\
The statistics of the polarization field can be obtained by calculating the two point correlation function. Similar to temperature, BipoSH coefficients can be also generalised to include the CMB polarization \cite{tsb}. The two point correlation between $E$, $B$ and $T$ can be defined as,
\begin{align}\label{eqbi3}
\bigg\langle X_{l_1m_1} X^{'*}_{l_2m_2}\bigg\rangle= \sum_{LM} A^{LM}_{l_1l_2|XX'}(-1)^{m_2}C^{LM}_{l_1m_1l_2-m_2} ,
\end{align}
where $X$ can be any one of $E$, $B$ and $T$. Here, similar to the temperature $A^{LM}_{l_1l_2|XX'}$, are the BipoSH coefficients for polarization. Under the assumption of  SI, the covariance matrix in the SH space is diagonal, which implies only $A^{00}_{ll|XX'} \neq 0$. Measurement of non-zero BipoSH values implies violation of SI.
\subsection{Minimum variance estimator}\label{app3}
We discuss here the reconstruction noise for the minimum variance estimator of the BipoSH coefficients of CMB polarization map. Minimum variance estimator is used for lensing reconstruction  by Hu \& Okamoto \cite{hu2}, Hanson et al. \cite{duncan, duncan2}. It is  also used by Planck for estimating several effects like lensing, SI violation and measuring $\beta$ from temperature map. \\ BipoSH coefficients for temperature and polarization of a CMB field, can be written as,
\begin{align}\label{eqnc1}
\hat{A}^{LM}_{ll'|XX'}   = {A}^{LM}_{ll'|XX'} + \alpha_{LM}S^{L}_{ll'|XX'}\, ,
\end{align}
where, $\hat{A}^{LM}_{ll'|XX'}$ is the observed BipoSH coefficient and ${A}^{LM}_{ll'|XX'}$ is the BipoSH coefficient for a SI temperature or polarization field, which on averaging over an ensemble is zero for $L \neq 0$. $\alpha_{LM}S^{L}_{ll'|XX'}$ is the source of SI violation with $S^{L}_{ll'|XX'}$ as the shape factor and $\alpha_{LM}$, strength of SI violation  arising due to inevitable effects like weak lensing, Doppler boost etc. For weak lensing, $\alpha_{LM}$ is equivalent to the lensing potential, $\phi_{LM}$, and for Doppler boost, $\alpha_{LM}$ is equivalent to local velocity $\beta_{1M}$  defined as \cite{Planck1, amendola},
\begin{align}\label{beta_def}
\begin{split}
\beta_{1M}&= \int \boldsymbol{\beta}. \hat n\,\, Y^{*}_{1M}(\hat n) d\hat n\, .\\
\end{split}
\end{align}
For the single CMB sky, ${A}^{1M}_{ll'|XX'}$  is non-zero and to measure the $\beta_{1M}$ , we define estimator, $\hat \beta_{LM}$ for $L=1$ as,
\begin{align}\label{eqnc2a}
\hat \beta_{1M}  = \sum_{ll'} \frac{{A}^{1M}_{ll'|XX'}}{S^{1}_{ll'|XX'}} + \beta_{1M}\, .
\end{align}
To estimate the significance of  signal, $\beta$ from the estimator, $\hat \beta$, we need to evaluate the variance of the estimator for SI sky. Detection of any statistically significant signal is only possible when variance of the first term in R.H.S. of eq.\eqref{eqnc2a} is smaller than $|\beta|^2$. Taking the product of eq.\eqref{eqnc2a}, with its complex conjugate, we get,
\begin{align}\label{eqnc3}
\begin{split}
(\hat \sigma_{  \beta})^2   \equiv \bigg\langle \hat \beta^*_{1M}\hat \beta_{1M'} \bigg\rangle  &= \sum_{l_1l_2l_3l_4} \frac{\bigg\langle {A}^{*1M}_{l_3l_4|XX'}{A}^{1M'}_{l_1l_2|XX'}\bigg\rangle}{S^{1}_{l_1l_2|XX'}S^{1}_{l_3l_4|XX'}} + |\beta_{1}|^2\, , \\
\,\\
(\hat \sigma_{ \beta})^2 &= N_{1} + |\beta_{1}|^{2},
\end{split}
\end{align}
where,\\
\begin{align}
N_{1}= \sum_{l_1l_2l_3l_4} \frac{\bigg\langle {A}^{*1M}_{l_3l_4|XX'}{A}^{1M'}_{l_1l_2|XX'}\bigg\rangle}{S^{1}_{l_1l_2|XX'}S^{1}_{l_3l_4|XX'}}.
\end{align}

$N_1$ is the reconstruction noise. 
In this work we first compute the reconstruction noise for CMB polarization map for BipoSH coefficients. The expressions for $N_1$ can be obtained by  using the variance of BipoSH coefficients for CMB polarization sky, computed in Sec.\ref{bipstat} . But due to weak signal strength, it is important to minimize the value of $N_1$ further. To arrive at the minimum variance estimator, we rewrite eq.\eqref{eqnc2a} as,
\begin{align}\label{eqnc4}
\hat \beta_{1M}  = \sum_{ll'}w^1_{ll'} \frac{{A}^{1M}_{ll'|XX'}}{S^{1}_{ll'|XX'}} + \beta_{1M}\, ,
\end{align}
where $w^1_{ll'}$ are the weights, which satisfies the constraint $\sum_{ll'} w^1_{ll'} =1$. This weight factors should be chosen such that, it minimizes the reconstruction noise. The variance of the eq.\eqref{eqnc4} is 
\begin{align}\label{eqnc5}
\begin{split}
(\hat \sigma_{ \beta})^2  &= \sum_{ll'} (w^1_{ll'})^2 \frac{\bigg\langle {A}^{*1M}_{ll'|XX'}{A}^{1M'}_{ll'|XX'}\bigg\rangle}{S^{1}_{ll'|XX'}S^{1}_{ll'|XX'}} + |\beta_{1}|^2\, .
\end{split}
\end{align}
Using the Lagrange multiplier method, we obtain the weight factors $w^1_{ll'}$ as,
\begin{align}\label{eqnc6}
w^1_{ll'}= \Bigg(\frac{(S^1_{ll'|XX'})^2}{\langle {A}^{*1M}_{ll'|XX'}{A}^{1M'}_{ll'|XX'}\rangle}\Bigg)\Bigg[\sum_{ll'} \frac{(S^1_{ll'|XX'})^2}{\langle {A}^{*1M}_{ll'|XX'}{A}^{1M'}_{ll'|XX'}\rangle} \Bigg]^{-1},
\end{align}
that minimizes the reconstruction noise $N_1$. Putting the value of  weight factor in eq.\eqref{eqnc5}, we get,
\begin{align}\label{eqnc7}
\begin{split}
(\hat \sigma_{\beta})^2  &= \bigg[\sum_{ll'}\frac{S^{1}_{ll'|XX'}S^{1}_{ll'|XX'}}{\langle {A}^{*1M}_{ll'|XX'}{A}^{1M'}_{ll'|XX'}\rangle}\bigg]^{-1} + |\beta_{1}|^2\, ,\\
\,\,
(\hat \sigma_{ \beta})^2  &= N_1 + |\beta_{1}|^2\,,
\end{split}
\end{align}
where, reconstruction noise is defined as, 
\begin{align}\label{eqnc8}
N_1= \bigg[\sum_{ll'}\frac{S^{1}_{ll'|XX'}S^{1}_{ll'|XX'}}{\langle {A}^{*1M}_{ll'|XX'}{A}^{1M'}_{ll'|XX'}\rangle}\bigg]^{-1},
\end{align}
and the minimum variance estimator is,
\begin{align}\label{eqnc9}
\hat \beta_{1M}  = \sum_{ll'}\frac{S^1_{ll'}\hat {A}^{1M}_{ll'|XX'}}{\bigg\langle {A}^{*1M}_{ll'|XX'}{A}^{1M'}_{ll'|XX'}\bigg\rangle}N_{1}  \, .
\end{align}
The product of inverse-variance filtered temperature map can be expressed in terms of the BipoSH coefficients and  variance of BipoSH coefficients. 
Minimum variance estimator derived here in terms of the BipoSH coefficients is equivalent to the estimator used by Planck \cite{Planck1} which is also discussed by Hanson \& Lewis \cite{duncan2}. A similar estimator has also been derived by Amendola et al. \cite{amendola} for Doppler boost estimation. In Sec.~\ref{noi_es} we use the above expression to infer the prospects of detecting and measuring the Doppler boost effect from polarization maps of Planck and future proposed mission PRISM.
\section{Effect of boost on CMB covariance matrix}
In the previous section, we have discussed the power spectra and bipolar representation of temperature and polarization field. The angular power spectra are the key prediction of the isotropic Cosmological model. So, to determine the correct Cosmological model it is necessary to account for the  Lorentz transformation of $C_l$ from CMB rest frame to the frame of the moving observer.  As the measured value of $\beta \approx10^{-3}$, we calculate the only leading order effect of Doppler boost on $C_l$. From the calculated Lorentz transformation of Stokes parameters in eq.\eqref{eq11} for small values of $\beta$, we can relate ${}_{\pm}\tilde{X}$ in the moving frame with ${}_{\pm}X$  in CMB rest frame as,
\begin{align}\label{eq19}
_\pm \tilde{X}(\hat n)= \, _{\pm}X\bigg(\hat n- \bigtriangledown(\hat n. \hat \beta)\bigg)\bigg(1+b_\nu\beta\cos\theta\bigg).
\end{align}
We notationally represent Doppler boosted quantities with over tilde. The Taylor series expansion of eq.\eqref{eq19} yields,
\begin{align}\label{eq20}
_\pm \tilde{X}(\hat n)= \bigg(1+b_\nu\beta\cos\theta\bigg)\bigg({}_\pm X(\hat n)- \bigtriangledown_i {}_\pm X(\hat n)\bigtriangledown^i(\hat n. \hat \beta) +\frac{1}{2}\bigtriangledown_j\bigtriangledown_i {}_\pm X(\hat n)\bigtriangledown^i(\hat n. \hat \beta)\bigtriangledown^j(\hat n. \hat \beta) \bigg),
\end{align}
where the covariant derivatives on SWSH are taken as defined in \cite{hu_2, challinor_2}. Here, the geometrical interpretation of this mapping is that one should parallel transport the Stokes parameters along a geodesic generated at $\hat n$ along the unit tangent vector, $\bigtriangledown^i(\hat n. \hat \beta)$.  Defining $\boldsymbol \beta$ in SH basis as mentioned in eq. \eqref{beta_def}, we rewrite eq.\eqref{eq20} in SWSH retaining terms up to  in $O(\beta^2)$  in aberration and $O(\beta)$ in modulation,
\begin{align}\label{eq22}
\begin{split}
_\pm \tilde{X}_{l_1m_1} = & {}_\pm X_{l_1m_1}- \sum_{l_2m_2}\sum_{m_3}\int d \hat n \beta_{1m_3} {}_\pm X_{l_2m_2} {}_{\pm2}Y_{l_1m_1}^*(\hat n) \bigtriangledown_i {}_{\pm2}Y_{l_2m_2}(\hat n)  \bigtriangledown^i Y_{1m_3}(\hat n) \\ &+\frac{1}{2}\sum_{l_2m_2}\sum_{m_3m_4}\int d \hat n {}_{\pm2}\beta_{1m_4}\beta_{1m_3} {}_\pm X_{l_2m_2}Y_{l_1m_1}^*(\hat n) \bigtriangledown_j\bigtriangledown_i {}_{\pm2}Y_{l_2m_2}(\hat n)  \bigtriangledown^i Y_{1m_3}(\hat n) \bigtriangledown^j Y_{1m_4}(\hat n)\\& +b_\nu\sum_M\sum_{l_2m_2}\int d \hat n \beta_{1M} {}_\pm X_{l_2m_2} Y_{1M}(\hat n) {}_{\pm2}Y_{l_1m_1}^*(\hat n) {}_{\pm2}Y_{l_2m_2}(\hat n)\\ &-b_\nu\sum_{Mm_3}\sum_{l_2m_2}\int d \hat n \beta_{1M}\beta_{1m_3} {}_\pm X_{l_2m_2}Y_{1M}(\hat n) {}_{\pm2}Y_{l_1m_1}^*(\hat n)\bigtriangledown_i {}_{\pm2}Y_{l_2m_2}(\hat n)\bigtriangledown^i Y_{1m_3}^*(\hat n).
\end{split}
\end{align}

Further, we define the quantities,
\begin{align}\label{eq23}
\begin{split}
_{\pm2}G_{l_11l_21}^{m_1Mm_2m_3}=& \int d \hat n\, Y_{1M}(\hat n) {}_{\pm2}Y_{l_1m_1}^*(\hat n)\bigtriangledown_i {}_{\pm2}Y_{l_2m_2}(\hat n) \bigtriangledown^i Y_{1m_3}(\hat n),\\
_{\pm2}H_{l_1l_21}^{m_1m_2M}=& \int d \hat n\, Y_{1M}(\hat n) {}_{\pm2}Y_{l_1m_1}^*(\hat n) {}_{\pm2}Y_{l_2m_2}(\hat n),
\end{split}
\end{align}
and as defined in \cite{wayne},
\begin{align}\label{eq24}
\begin{split}
_{\pm2}I_{l_1l_21}^{m_1m_2m_3}=& \int d \hat n\, \, {}_{\pm2}Y_{l_1m_1}^*(\hat n) \bigtriangledown_i {}_{\pm2}Y_{l_2m_2}(\hat n)\bigtriangledown^i Y_{1m_3}(\hat n),\\
_{\pm2}J_{l_11l_21}^{m_1m_3m_2m_4}=& \int d \hat n\, \, {}_{\pm2}Y_{l_1m_1}^*(\hat n) \bigtriangledown_j\bigtriangledown_i \,\,{}_{\pm2}Y_{l_2m_2}(\hat n) \bigtriangledown^i Y_{1m_3}(\hat n)\bigtriangledown^j Y_{1m_4}^*(\hat n).
\end{split}
\end{align}
With these definitions, eq.\eqref{eq22} becomes,
\begin{align}\label{eq25}
\begin{split}
_\pm \tilde{X}_{l_1m_1} =& {}_\pm X_{l_1m_1}- 
\sum_{l_2m_2}\sum_{m_3} {}_\pm X_{l_2m_2}
\beta_{1m_3} {}_{\pm2}I_{l_1l_21}^{m_1m_2m_3}
+\frac{1}{2}\sum_{l_2m_2}\sum_{m_3m_4} {}_\pm X_{l_2m_2}
\beta_{1m_3} \beta_{1m_4} {}_{\pm2}J_{l_11l_21}^{m_1m_3m_2m_4}\\& +b_\nu\sum_M\sum_{l_2m_2} {}_\pm 
X_{l_2m_2} \beta_{1M} {}_{\pm2}H_{l_1l_21}^{m_1m_2M}-b_\nu\sum_{Mm_3}\sum_{l_2m_2}  {}_\pm X_{l_2m_2}\beta_{1M} \beta_{1m_3}{}_{\pm2}G_{l_11l_21}^{m_1Mm_2m_3}.
\end{split}
\end{align}
The power spectrum is obtained by taking the product of eq.\eqref{eq25} with its complex conjugate. We calculate the  effect of Lorentz transformation on both diagonal and off-diagonal terms of SH space covariance matrix. First, the effect on diagonal part (angular power spectra), followed by the effect on off-diagonal terms (BipoSH coefficients) are calculated in the next two subsections.

\subsection{Effect  of boost on angular power spectra $C_l^{XX'}$}\label{po2}
In this section, we calculate the effect of Doppler boost on angular power spectra. The Lorentz transformation of these diagonal terms of SH covariance matrix can be calculated by taking the product of $\langle {}_\pm \tilde{X}^*_{l_1m_1}{}_\pm \tilde{X}_{l_2m_2}\rangle$ and using the conditions given in eq.\eqref{eq18}. Since $E$ and $B$ are uncorrelated for different $l$ values, angular power spectra is not affected by the terms at $O(\beta)$.\\ The leading order  effect on the  diagonal elements are O($\beta^2$). We take the product $_- \tilde{X}_{l_1m_2}$ with $_-\tilde{X}^*_{l_2m_2}$ and $_+ \tilde{X}^*_{l_2m_2}$, and retaining only the  $O(\beta^2)$ terms, gives us the leading  O($\beta^2$) correction terms. In  Appendix \ref{app2}, we have provided details of the intermediate steps of the calculation. The correction to the angular power spectrum are,

\begin{align}\label{eq44}
\begin{split}
\tilde{C}^{EE}_{l} =& \, \,C^{EE}_{l}\bigg[1+ \beta^2 \left({}_{s}R_l -2b_\nu \, \, _{2}Q^{m\, m\,M\, M}_{l \, l \,1 \,1}\right)\bigg] + \beta^2(b_\nu -1)^2 C^{BB}_{l} {_{2}}M^2_{l \, l \,1}\\& + \beta^2 C^{EE}_{l +1}\, _{2}Z_{l+1 \, l\,1}^2 +  \beta^2 C^{EE}_{l -1}\, _{2}Z_{l-1\, l\,1}^2,
 \end{split}
\end{align}
\begin{align}\label{eq44a}
\begin{split}
\tilde{C}^{BB}_{l} =& \, \,C^{BB}_{l}\bigg[1+ \beta^2 \left ({}_{s}R_l -2b_\nu \, \, _{2}Q^{m\,\,M\,M}_{l \,l \,1 \,1}\right)\bigg] + \beta^2(b_\nu -1)^2 C^{EE}_{l} {_{2}}M^2_{l \,l \,1}\\& + \beta^2 C^{BB}_{l +1}\, _{2}Z_{l+1\,l\,1}^2 +  \beta^2 C^{BB}_{l -1}\, _{2}Z_{l-1\,l\,1}^2,
 \end{split}
\end{align}

\begin{align}\label{eq43b}
\begin{split}
\tilde{C}^{TE}_{l} =& \, \,C^{TE}_{l}\bigg[1+ \frac{\beta^2}{2} \bigg(\bigg({}_{0}R_l+{}_{s}R_l\bigg) - 2b_\nu{}\bigg( {}_{2}Q^{m\,m\,M\,M}_{l \, l\, 1 \,1} + \, _{0}Q^{m\,m\,M\,M}_{l\, l \,1 \,1}\bigg) \bigg)\bigg]\\& + \beta^2 C^{TE}_{l +1}\,\, _{2}Z_{l+1\,l\,1}\, \, _{0}Z_{l+1\,l\,1} +  \beta^2 C^{TE}_{l -1}\,\, _{2}Z_{l-1\,l\,1}\, \, _{0}Z_{l-1\,l\,1},
 \end{split}
\end{align}


where, we define,\\
\begin{align}\label{eq28c1}
\begin{split}
& _{\pm s}M_{l'\,l \,1}= \frac{\Pi_{l'}}{\sqrt{4\pi}} C^{10}_{l'\, s \,l\, -s},\\
&_{\pm s}N_{l+1\,l \,1}= \frac{\Pi_{l+1}}{\sqrt {4\pi}} (l+2)C^{10}_{l+1\, s \,l \,-s},\\
&_{\pm s}N_{l-1\,l \,1}= \frac{\Pi_{l-1}}{\sqrt {4\pi}} (l-1)C^{10}_{l-1 \,s\, l\, -s}.\\
& {}_{\pm s}Q^{m\, m\, M\, M}_{l\, l\, 1\, 1} \equiv \, \sum_{M} {}_{\pm s}G^{m\,m\,M\,M}_{l\,l \,1 \,1}  = \sum_{J} \frac{\Pi_{J}^2\bigg[2 +l(l+1) - J(J+1)\bigg]}{8\pi} C^{1 0}_{l\, s \,J \,-s}C^{1 0}_{l \,s \,J \,- s} ,\\
&{}_{\pm s}R_l= \frac{3(l(l-1)-s^2)}{4\pi},\\
&\, {}_{s}Z_{l'\, l\, 1}= \, ({}_{s}N_{l'\,l \,1} - \, b_\nu \, {}_{s}M_{l'\,l \,1}),\\
\end{split}
\end{align}
with the notation $\Pi_{l_1\, l_2 \ldots l_n}= \sqrt{(2l_1+1)(2l_2+1)\ldots (2l_n+1)}$ \cite{varsha}.\\
The eqs.\eqref{eq44}, \eqref{eq44a} and \eqref{eq43b}  give the correction due to the Lorentz transformation on the diagonal elements of $EE$, $BB$ and $TE$.  Effect of Doppler boost on $C_l$ has been also computed by Challinor \& Leeuwen \cite{challinor} for a frequency integrated $X_{lm}$. Our results in eqs.\eqref{eq44}, \eqref{eq44a} and \eqref{eq43b} are given in terms of BipoSH coefficients including $b_{\nu}$ factor in the modulation term, otherwise these expressions can also be obtained from \cite{challinor}. In the angular power spectra there is a mild mixing between $EE$ and $BB$ polarization.   In the next section we derive the expression of BipoSH coefficients assuming SI in CMB rest frame and show that the effects appear at linear order in $\beta$.
\subsection{Effect of boost on BipoSH spectra ${A}^{LM}_{ll' | _\pm X _\pm X}$}
As we have mentioned earlier, the two point correlation function of the CMB temperature anisotropy can be expressed as 
\begin{equation}\label{eq31a}
\bigg\langle {}_\pm \tilde{X}^*_{l_1m_1} {}_\pm \tilde{X}_{l_2m_2}\bigg\rangle= \sum_{LM} {\tilde{A}^{LM}_{l_1l_2| {}_\pm X {}_\pm X}(-1)^{m_1}C^{LM}_{l_1-m_1l_2m_2}}.
\end{equation}
where, we define $\tilde{A}^{LM}_{l_1l_2 | _\pm X _\pm X}$ are the BipoSH coefficients. These completely encode the off-diagonal elements of the SH space covariance matrix and non zero value of these coefficients are the signature of SI violation. The BipoSH coefficients can be decomposed into real ($^R\tilde{A}^{LM}_{ll' | _\pm X _\pm X}$) and imaginary ($^I\tilde{A}^{LM}_{l_1l_2 | _\pm X _\pm X}$) parts,\\
\begin{equation}\label{eqr1a}
\tilde{A}^{LM}_{l_1l_2 | _\pm X _\pm X}= {}^R\tilde{A}^{LM}_{l_1l_2 | _\pm X _\pm X}+ i \,\,{}^I\tilde{A}^{LM}_{l_1l_2 | _\pm X _\pm X}.
\end{equation}
Now using the property,
\begin{equation}\label{eqr1}
{}_\pm \tilde{X}^*_{l\,m}= (-1)^m{}_\mp \tilde{X}_{l\,-m},
\end{equation}
the product of   eq.\eqref{eqr1} with its complex conjugate leads to, 
 \begin{equation}\label{eqr2}
 \bigg\langle {}_\pm \tilde{X}^*_{l_1m_1} {}_\pm \tilde{X}_{l_2m_2}\bigg\rangle^*= (-1)^{m_1+m_2}\bigg \langle {}_\mp \tilde{X}^*_{l_1\,-m_1} {}_\mp \tilde{X}_{l_2 \,-m_2}\bigg\rangle .
\end{equation}
This implies the $EE$, $BB$ and $TE$ correlations are the real part of the BipoSH coefficients, which we can express for $L=1,M= 0$ as,
\begin{align}\label{eq48}
\begin{split}
2\tilde{A}^{10}_{ll+1|EE}&= \sum_{mm'} \bigg[\bigg\langle {}_+\tilde{X}_{lm}^* \,{}_+ \tilde{X}_{l+1m'}\bigg\rangle + \bigg\langle {}_-\tilde{X}_{lm}^*\,{}_+ \tilde{X}_{l+1m'}\bigg\rangle\bigg] (-1)^{m}C^{10}_{l\,-m\,l+1\,m'},\\
2\tilde{A}^{10}_{ll+1|BB}&= \sum_{mm'} \bigg[\bigg\langle {}_+\tilde{X}_{lm}^*\, {}_+ \tilde{X}_{l+1m'}\bigg\rangle - \bigg\langle {}_-\tilde{X}_{lm}^* \,{}_+ \tilde{X}_{l+1m'}\bigg\rangle\bigg] (-1)^{m}C^{10}_{l\,-m\,l+1\,m'},\\
2\tilde{A}^{10}_{ll+1|TE}&= \sum_{mm'}\bigg[\bigg \langle {}_+\tilde{X}_{lm}^*\, \tilde a_{l+1m'}\bigg\rangle + \bigg\langle {}_-\tilde{X}_{lm}^*\, \tilde a_{l+1m'}\bigg\rangle\bigg] (-1)^{m}C^{10}_{l\,-m\,l+1m'},
\end{split}
\end{align}
and the imaginary parts of the covariance matrix are $EB$ and $TB$, which are related to corresponding BipoSH coefficients for $L=1,M= 0$ by,
\begin{align}\label{eq48ra}
\begin{split}
\tilde{A}^{10}_{ll|EB}&=  \sum_{mm'} \bigg\langle{E}_{lm}^* {B}_{lm'}\bigg \rangle (-1)^{m}C^{10}_{l\,-mlm'}\\& =\frac{i}{2} \sum_{mm'} \bigg[\bigg\langle {}_+\tilde{X}_{lm}^* {}_+\tilde{X}_{lm'} \bigg\rangle + \bigg\langle {}_-\tilde{X}_{lm}^* {}_+\tilde{X}_{lm'}\bigg\rangle\bigg] (-1)^{m}C^{10}_{l\,-mlm'}, \\
\tilde{A}^{10}_{ll|TB}&=  \sum_{mm'} \bigg\langle {B}_{lm}^*{a}_{lm'}\bigg\rangle (-1)^{m}C^{10}_{l\,-mlm'} \\&= \frac{i}{2}\sum_{mm'} \bigg[\bigg\langle {}_+\tilde{X}_{lm}^*\tilde a_{lm'} \bigg\rangle - \bigg\langle {}_-\tilde{X}_{lm}^* \tilde a_{lm'}\bigg\rangle \bigg] (-1)^{m}C^{10}_{l\,-mlm'}.
\end{split}
\end{align}
For simplicity we choose the direction of $\boldsymbol \beta$ along $\hat z$ direction, which reduces eq.~\eqref{beta_def} to\\
\begin{align}\label{eqb1}
\beta_{1M}\equiv 2\sqrt{\frac{\pi}{3}}\beta \delta_{M0} = 2.52 \times 10^{-3} .
\end{align}
The effect in any other sky coordinate can be recovered by rotation transformation of BipoSH coefficients. The non-zero BipoSH coefficients arising due to Lorentz transformation of Stokes parameters from CMB rest frame to the moving observer frame are,

\begin{align}\label{eq40}
\begin{split}
\bigg\langle {}_-\tilde{X}_{l_1m_1}^* {}_+ \tilde{X}_{l_2m_2}\bigg\rangle =&\beta_{10} \bigg[-\frac{1}{2}(C^{EE}_{l_2} - C^{BB}_{l_2})  \big[l_2(l_2+1)+2-l_1(l_1+1)\big](-1)^{l_1+l_2+1}  \\& + b_\nu(C^{EE}_{l_2} - C^{BB}_{l_2}) (-1)^{l_1+l_2+1}\\&  -\frac{1}{2}(C^{EE}_{l_1} - C^{BB}_{l_1})[l_1(l_1+1)+2-l_2(l_2+1)]\\& + b_\nu(C^{EE}_{l_1} - C^{BB}_{l_1}) \bigg ] (-1)^{m_1}\frac{\Pi_{l_1\,l_2}}{\sqrt{4\pi}\Pi_{1}}C^{1 0}_{l_1 2 l_2 -2}C^{10}_{l_1-m_1l_2m_2},\\
 \end{split}
\end{align}

\begin{align}\label{eq41}
\begin{split}
\bigg\langle {}_+\tilde{X}_{l_1m_1}^* {}_+ \tilde{X}_{l_2m_2}\bigg\rangle = &\beta_{10} \bigg[-\frac{1}{2}(C^{EE}_{l_2} + C^{BB}_{l_2})  [l_2(l_2+1)+2-l_1(l_1+1)]  \\& +  b_\nu(C^{EE}_{l_2} + C^{BB}_{l_2})   \\&  - \frac{1}{2}(C^{EE}_{l_1} + C^{BB}_{l_1})  [l_1(l_1+1)+2-l_2(l_2+1)] \\& + b_\nu(C^{EE}_{l_1} + C^{BB}_{l_1}) \bigg](-1)^{m_1}\frac{\Pi_{l_1\, l_2}}{\sqrt{4\pi}\Pi_{1}}C^{1 0}_{l_1 2 l_2 -2}C^{10}_{l_1-m_1l_2m_2}.\\
\end{split}
\end{align}

\begin{align}\label{eq41a}
\begin{split}
\bigg\langle {}_+\tilde{X}_{l_1m_1}^*\tilde a_{l_2m_2}\bigg\rangle =  \beta_{10} \bigg[&-\frac{1}{2}C^{TE}_{l_2}  [l_2(l_2+1)+2-l_1(l_1+1)]\frac{\Pi_{l_1\, l_2}}{\sqrt{4\pi}\Pi_{1}}C^{1 0}_{l_2  2 l_1 -2}  \\& +  b_\nu C^{TE}_{l_2}  \frac{\Pi_{l_1\, l_2}}{\sqrt{4\pi}\Pi_{1}}C^{1 0}_{l_2 2 l_1 -2} \\&  - \frac{1}{2}C^{TE}_{l_1}   [l_1(l_1+1)+2-l_2(l_2+1)]\frac{\Pi_{l_1\, l_2}}{\sqrt{4\pi}\Pi_{1}}C^{1 0}_{l_1  0 l_2 0} \\& + b_\nu C^{TE}_{l_1}  \frac{\Pi_{l_1\, l_2}}{\sqrt{4\pi}\Pi_{1}}C^{1 0}_{l_1  0 l_2 0}\bigg](-1)^{m_1}C^{10}_{l_1-m_1l_2m_2}.\\
\end{split}
\end{align}
BipoSH coefficients can be decomposed into even ($^{+}A^{LM}_{ll'}$) and odd ($^{-}A^{LM}_{ll'}$) parity \cite{bookta}. By decomposing the BipoSH coefficients for polarization into odd and even parity, we can write,
\begin{align}\label{eqp1}
A^{LM}_{ll'|{}_{\pm}X{}_{\pm}X'}= \,^{+}A^{+LM}_{ll'|{}_{\pm}X{}_{\pm}X'} \bigg[\frac{1+ (-1)^{l+l'+L}}{2}\bigg] + \, ^{-}A^{LM}_{ll'|{}_{\pm}X{}_{\pm}X'} \bigg[\frac{1- (-1)^{l+l'+L}}{2}\bigg].
\end{align}
Even (odd) parity BipoSH are zero for the value of the sum $l+l'+L$ being odd (even). Also, even (odd) parity BipoSH are symmetric (antisymmetric) in $l$ and $l'$. Since eqs.\eqref{eq40} and \eqref{eq41} are symmetric in $l_1$ and $l_2$ for $l_2= l_1\pm1$, these terms are the even parity BipoSH coefficients. But for $l_2 =l_1$, we have odd parity BipoSH coefficients.  Using eq.\eqref{eq48} we obtain even BipoSH coefficients  the $EE$, $BB$ and $TE$ from eqs.\eqref{eq40} and \eqref{eq41}as,

\begin{align}\label{eq49}
\begin{split}
\tilde{A}^{10}_{ll+1|E E}   =& \beta_{10} \bigg[C^{EE}_{l}(l+ b_\nu) -(l+2- b_\nu)C^{EE}_{l+1} \bigg]\frac{\Pi_{l\,l+1}}{\sqrt{4\pi}\Pi_{1}}C^{1 0}_{l+1\,- 2\, l\, 2},\\
 =& \, 2\pi \, \frac{D^{EE}_l }{l(l+1)}\frac{\Pi_{l\,l+1}}{\sqrt{4\pi}\Pi_{1}}C^{1 0}_{l+1\,- 2\, l\, 2},
\end{split}
\end{align}

\begin{align}\label{eq50}
\begin{split}
\tilde{A}^{10}_{ll+1|BB}   =& \beta_{10} \bigg[C^{BB}_{l}(l+ b_\nu) -(l+2-  b_\nu)C^{BB}_{l+1}\bigg]\frac{\Pi_{l\,l+1}}{\sqrt{4\pi}\Pi_{1}}C^{1 0}_{l+1\,- 2\, l\, 2},\\
 =& \, 2\pi \, \frac{D^{BB}_l }{l(l+1)}\frac{\Pi_{l\,l+1}}{\sqrt{4\pi}\Pi_{1}}C^{1 0}_{l+1\,- 2\, l\, 2},
\end{split}
\end{align}
where, we define BipoSH spectra for $Y=E, B$ polarization as,
\begin{align}\label{eqd2}
\begin{split}
D^{YY}_l =\, \frac{l(l+1)\beta_{10} \bigg[(l+ b_\nu)C^{YY}_{l} -(l+2- b_\nu)C^{YY}_{l+1}\bigg]}{2\pi}.
\end{split}
\end{align}
The cross correlation between $T$ and $E$ can be obtained as,

\begin{align}\label{eq51}
\begin{split}
\tilde{A}^{10}_{ll+1|TE}   =& \beta_{10} \bigg [(l+ b_\nu)C^{TE}_{l} C^{1 0}_{l \,0 \,l+1 \,0}-(l+2- b_\nu) C^{TE}_{l+1}C^{1 0}_{l\, -2\, l+1\, 2}\bigg]\frac{\Pi_{ll+1}}{\sqrt{4\pi}\Pi_{1}},\\
=& \, \, 2\pi \frac{D^{TE}_{l}}{l(l+1)}\frac{\Pi_{ll+1}}{\sqrt{4\pi}\Pi_{1}}C^{1 0}_{l\, -2\, l+1\, 2},
\end{split}
\end{align}
where, we define BipoSH spectra for $TE$ correlation as,
\begin{align}\label{eqd2}
\begin{split}
D^{TE}_l= \, \frac{\beta_{10} l(l+1)\bigg[(l+ b_\nu)C^{TE}_{l}\frac{C^{1 0}_{l \,0 \,l+1 \,0}}{C^{1 0}_{l\, -2\, l+1\, 2}} -(l+2 -b_\nu)C^{TE}_{l+1}\bigg]}{{2\pi}} .
\end{split}
\end{align}
Similar result for $A_{ll+1|TT}^{10}$ is obtained \cite{Planck1},
\begin{align}\label{eqd3}
\begin{split}
\tilde{A}^{10}_{ll+1|TT}= 2\pi \frac{D^{TT}_{l}}{l(l+1)}\frac{\Pi_{ll+1}}{\sqrt{4\pi}\Pi_{1}}C^{1 0}_{l\, 0\, l+1\, 0},
\end{split}
\end{align}
where,
\begin{align}\label{eqd3}
\begin{split}
D^{TT}_l= \, \frac{\beta_{10} l(l+1)\bigg[(l+ b_\nu)C^{TT}_{l} -(l+2 -b_\nu)C^{TT}_{l+1}\bigg]}{{2\pi}} .
\end{split}
\end{align}
From the above expression it is clear that there is non zero BipoSH coefficient only for $L=1$ arising from the first order terms in $\beta$. It is also interesting to point out that there is no mixing of $EE$ and $BB$ polarization in BipoSH coefficients due to Doppler boost. This is because $E$ and $B$ are uncorrelated for a SI CMB polarization field.  
Similar to the even parity BipoSH coefficients there are also odd parity BipoSH coefficients with $l_1=l_2$. The parity violating terms  are the cross-correlation between $EB$ and $TB$. This implies,
 
 \begin{align}\label{eq51a}
\begin{split}
 \tilde{A}^{10}_{ll|EB}   = i& \, 2\pi \, \frac{D^{EB}_l }{l(l+1)}\frac{\Pi^2_{l}}{\sqrt{4\pi}\Pi_{1}}C^{1 0}_{l\, -2\, l\, 2},\\
\tilde{A}^{10}_{ll|TB}   =& \, \, i2\pi \frac{D^{TB}_{l}}{l(l+1)}\frac{\Pi^2_{l}}{\sqrt{4\pi}\Pi_{1}}C^{1 0}_{l\, -2\, l\, 2},
\end{split}
\end{align}
where BipoSH spectra with odd parity  can be written as,

\begin{align}\label{eqd2a}
\begin{split}
D^{EB}_l&= \, \frac{\beta_{10} l(l+1)(C^{EE}_{l} + C_{l}^{BB})(b_\nu -1)}{{2\pi}},\\
D^{TB}_l&= \, \frac{\beta_{10} l(l+1)(b_\nu -1)C^{TE}_{l}}{{2\pi}} .
\end{split}
\end{align}

\begin{figure}[H]
\centering
\includegraphics[width=5.0in,keepaspectratio=true]{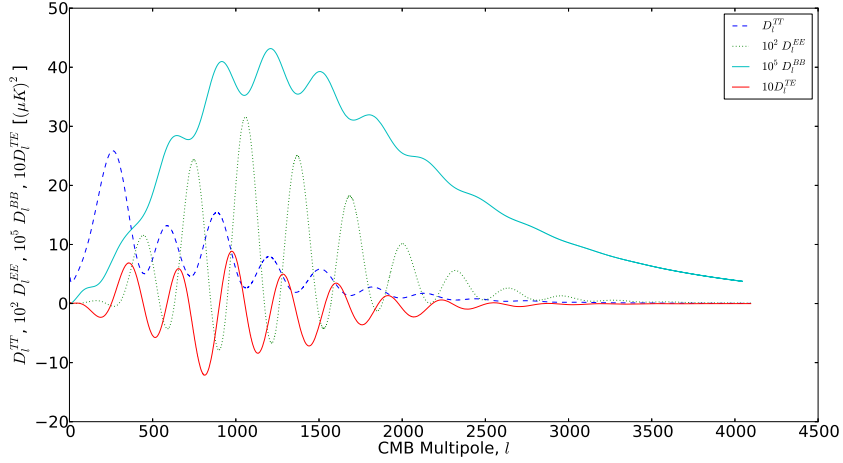}
\caption{BipoSH spectra, $D_l^{TT}$, $D_l^{EE}$, $D_l^{BB}$ and $D_l^{TE}$, with $b_\nu =3 $ for $\nu= 217$ GHz arising due to boost with $\beta= 1.23 \times10^{-3}$. Here we have used the best-fit $\Lambda CDM$ $C_l^{TT}, C_l^{EE}, C_l^{BB}$ and $C_l^{TE}$ generated from CAMB\cite{camb}.}\label{fig:e}
\end{figure}

\begin{figure}[H]
\centering
\includegraphics[width=5.0in,keepaspectratio=true]{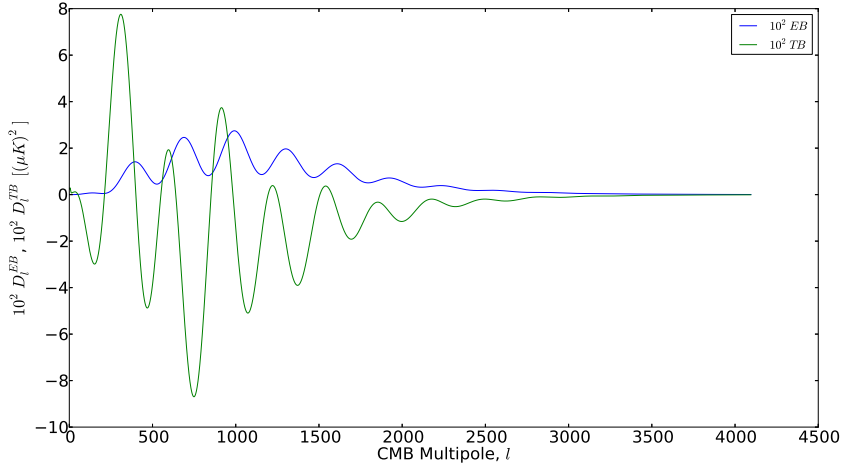}
\caption{BipoSH spectra, $D_l^{EB}$ and $D_l^{TB}$ with $b_\nu =3 $ for $\nu= 217$ GHz arising due to Doppler boost with $\beta= 1.23 \times10^{-3}$. Here we have used the best-fit $\Lambda CDM$ $C_l^{TT}$,$C_l^{EE}, C_l^{BB}$ and $C_l^{TE}$ generated from CAMB\cite{camb}.}\label{fig:e1}
\end{figure}

We compute BipoSH spectra in the above eqs.\eqref{eq49}, \eqref{eq50}, \eqref{eq51} using the lensed $C_l^{EE}$, $C_l^{BB}$ and $C_l^{TE}$ from CAMB \cite{camb} with the best fit $\Lambda CDM$ parameters from Planck \cite{Planck3}. We use the value of tensor scale ratio, $r$, as $0.1$. The estimated BipoSH spectra are computed and plotted in fig.\ref{fig:e} with $\beta= 1.23 \times10^{-3}$ \cite{WMAP} for frequency channel $\nu = 217\, \text{GHz}$, for which $b_\nu = 3$. The BipoSH spectra for $EB$ and $TB$ mentioned in eq.\eqref{eqd2a} are plotted in fig.\ref{fig:e1}. This also sets the inevitable bias in the measure of SI violation of CMB polarization while searching for Cosmological signature of SI violation. 

\section{Statistics of BipoSH coefficients for polarization}\label{bipstat}
In the previous section we estimate the BipoSH coefficients arising due to our local motion. However, to estimate $\beta$ from the BipoSH measurements, we need to study the statistics of the BipoSH coefficients for polarization. Statistics of diagonal elements for polarization are discussed by \cite{marc,kin,ftcmb}.
The statistics of the BipoSH coefficients, ${A}^{LM}_{ll'|XX'}$, can be obtained with the assumption that $\langle {A}^{LM}_{ll'|XX'}\rangle =0$. The variance of the BipoSH coefficients can  be written as,
\begin{align}\label{eqes1}
\begin{split}
\langle {A}^{LM}_{l_1l_2|EE}{A}^{L'M'*}_{l_3l_4|EE} \rangle = &\frac{1}{16}\bigg[\sum_{m_1m_2m_3m_4}  \bigg \langle \bigg( {}_+\tilde{X}^*_{l_3m_3}{}_+ \tilde{X}_{l_4m_4} +{}_-\tilde{X}^*_{l_3m_3}{}_- \tilde{X}_{l_4m_4} + {}_-\tilde{X}^*_{l_3m_3}{}_+ \tilde{X}_{l_4m_4} \\& +{}_+\tilde{X}^*_{l_3m_3}{}_- \tilde{X}_{l_4m_4} \bigg)^*\bigg({}_+\tilde{X}^*_{l_1m_1}{}_+ \tilde{X}_{l_2m_2}+ {}_-\tilde{X}^*_{l_1m_1}{}_- \tilde{X}_{l_2m_2}+ {}_+\tilde{X}^*_{l_1m_1}{}_- \tilde{X}_{l_2m_2}\\&+  {}_-\tilde{X}^*_{l_1m_1}{}_+ \tilde{X}_{l_2m_2}\bigg)\bigg \rangle C^{L'M'}_{{l_4} m_4 l_3 -m_3}C^{LM}_{{l_2} m_2 l_1 -m_1}\bigg],\\
\langle  {A}^{LM}_{l_1l_2|BB}{A}^{L'M'*}_{l_3l_4|BB}\rangle = &\frac{1}{16}\bigg[\sum_{m_1m_2m_3m_4}  \bigg \langle \bigg({}_+\tilde{X}^*_{l_3m_3}{}_+ \tilde{X}_{l_4m_4} +{}_-\tilde{X}^*_{l_3m_3}{}_- \tilde{X}_{l_4m_4} - {}_-\tilde{X}^*_{l_3m_3}{}_+ \tilde{X}_{l_4m_4} \\&-{}_+\tilde{X}^*_{l_3m_3}{}_- \tilde{X}_{l_4m_4} \bigg)^*  \bigg({}_+\tilde{X}^*_{l_1m_1}{}_+ \tilde{X}_{l_2m_2}+ {}_-\tilde{X}^*_{l_1m_1}{}_- \tilde{X}_{l_2m_2}- {}_+\tilde{X}^*_{l_1m_1}{}_- \tilde{X}_{l_2m_2} \\&- {}_-\tilde{X}^*_{l_1m_1}{}_+ \tilde{X}_{l_2m_2}\bigg)\bigg \rangle C^{L'M'}_{{l_4} m_4 l_3 -m_3}C^{LM}_{{l_2} m_2 l_1 -m_1}\bigg],\\
\langle {A}^{LM}_{l_1l_2|TE}{A}^{L'M'*}_{l_3l_4|TE} \rangle = & \frac{1}{4}\bigg[\sum_{m_1m_2m_3m_4}  \bigg \langle \bigg({}_+\tilde{X}_{l_3m_3}^* a_{l_4m_4} + {}_-\tilde{X}^*_{l_3m_3}a_{l_4m_4}\bigg)^* \bigg({}_+\tilde{X}^*_{l_1m_1}a_{l_2m_2}+ {}_-\tilde{X}^*_{l_1m_1}a_{l_2m_2}\bigg)\bigg \rangle \\&C^{L'M'}_{{l_4} m_4 l_3 -m_3}C^{LM}_{{l_2} m_2 l_1 -m_1}\bigg],\\
\langle {A}^{LM}_{l_1l_2|EB}{A}^{L'M'*}_{l_3l_4|EB} \rangle = &\frac{1}{16}\bigg[\sum_{m_1m_2m_3m_4}  \bigg \langle \bigg({}_+\tilde{X}^*_{l_3m_3}{}_+ \tilde{X}_{l_4m_4} -{}_-\tilde{X}^*_{l_3m_3}{}_- \tilde{X}_{l_4m_4} + {}_-\tilde{X}^*_{l_3m_3}{}_+ \tilde{X}_{l_4m_4} \\&- {}_+\tilde{X}^*_{l_3m_3}{}_- \tilde{X}_{l_4m_4} \bigg)^* \bigg({}_+\tilde{X}^*_{l_1m_1}{}_+ \tilde{X}_{l_2m_2}- {}_-\tilde{X}^*_{l_1m_1}{}_- \tilde{X}_{l_2m_2}- {}_+\tilde{X}^*_{l_1m_1}{}_- \tilde{X}_{l_2m_2} \\&+ {}_-\tilde{X}^*_{l_1m_1}{}_+ \tilde{X}_{l_2m_2}\bigg)\bigg \rangle  C^{L'M'}_{{l_4} m_4 l_3 -m_3}C^{LM}_{{l_2} m_2 l_1 -m_1}\bigg],\\
\langle {A}^{LM}_{l_1l_2|TB}{A}^{L'M'*}_{l_3l_4|TB} \rangle = &\frac{1}{4}\bigg[\sum_{m_1m_2m_3m_4}  \bigg \langle \bigg({}_+\tilde{X}^*_{l_3m_3}a_{l_4m_4} -{}_-\tilde{X}^*_{l_3m_3}a_{l_4m_4}\bigg)^* \bigg({}_+\tilde{X}^*_{l_1m_1}a_{l_2m_2}- {}_-\tilde{X}^*_{l_1m_1}a_{l_2m_2}\bigg)\bigg \rangle \\&C^{L'M'}_{{l_4} m_4 l_3 -m_3}C^{LM}_{{l_2} m_2 l_1 -m_1}\bigg],\\
\langle {A}^{LM}_{l_1l_2|TT}{A}^{L'M'*}_{l_3l_4|EE} \rangle = &\frac{1}{4}\bigg[\sum_{m_1m_2m_3m_4}  \bigg \langle \bigg({}_+\tilde{X}^*_{l_3m_3}{}_+\tilde{X}_{l_4m_4} +{}_-\tilde{X}^*_{l_3m_3}{}_+\tilde{X}_{l_4m_4} + {}_-\tilde{X}^*_{l_3m_3}{}_-\tilde{X}_{l_4m_4}\\& +{}_+\tilde{X}^*_{l_3m_3}{}_-\tilde{X}_{l_4m_4} \bigg)^* \bigg(a^*_{l_1m_1}a_{l_2m_2}\bigg)\bigg \rangle C^{L'M'}_{{l_4} m_4 l_3 -m_3}C^{LM}_{{l_2} m_2 l_1 -m_1}\bigg],\\
\langle {A}^{LM*}_{l_1l_2|TT}{A}^{L'M'}_{l_3l_4|TE} \rangle = &\frac{1}{2}\bigg[\sum_{m_1m_2m_3m_4}  \bigg \langle \bigg(a^*_{l_1m_1}a_{l_2m_2}\bigg)^* \bigg({}_+\tilde{X}_{l_3m_3}^* a_{l_4m_4} + {}_-\tilde{X}^*_{l_3m_3}a_{l_4m_4}\bigg)\bigg \rangle C^{L'M'}_{{l_4} m_4 l_3 -m_3}C^{LM}_{{l_2} m_2 l_1 -m_1}\bigg],\\
\langle {A}^{LM*}_{l_1l_2|TE}{A}^{L'M'}_{l_3l_4|EE} \rangle = &\frac{1}{8}\bigg[\sum_{m_1m_2m_3m_4}  \bigg \langle \bigg({}_+\tilde{X}^*_{l_1m_1}a_{l_2m_2}+ {}_-\tilde{X}^*_{l_1m_1}a_{l_2m_2}\bigg)^* \bigg({}_+\tilde{X}^*_{l_3m_3}{}_+\tilde{X}_{l_4m_4} +{}_-\tilde{X}^*_{l_3m_3}{}_+\tilde{X}_{l_4m_4} \\& + {}_-\tilde{X}^*_{l_3m_3}{}_-\tilde{X}_{l_4m_4} +{}_+\tilde{X}^*_{l_3m_3}{}_-\tilde{X}_{l_4m_4} \bigg) \bigg \rangle C^{L'M'}_{{l_4} m_4 l_3 -m_3}C^{LM}_{{l_2} m_2 l_1 -m_1}\bigg].
\end{split}
\end{align}
With the assumption that the random temperature fluctuations are Gaussian distribution, we can reduce the four point correlation function, $\langle {A}^{L'M'}_{l_1l_2|XX'} {A}^{LM}_{l_3l_4|XX'}\rangle$  as products of two point correlation functions. Applying this to eq.\eqref{eqes1}, we get the variance of BipoSH coefficients with $L \neq 0$ for polarization as ,
\begin{align}\label{eqes2}
\begin{split}
\bigg\langle {A}^{LM}_{l_1l_2|EE} {A}^{L'M'*}_{l_3l_4|EE} \bigg\rangle = \,\,&(-1)^{l_1+l_2+L} \,\,C_{l_1}^{EE}C_{l_2}^{EE}\delta_{LL'}\delta_{MM'}\delta_{l_1l_4}\delta_{l_3l_2}\\ & + C_{l_1}^{EE}C_{l_2}^{EE}\delta_{LL'}\delta_{MM'}\delta_{l_2l_4}\delta_{l_3l_1},\\
\,\\
\bigg\langle  {A}^{LM}_{l_1l_2|BB}{A}^{L'M'*}_{l_3l_4|BB}\bigg\rangle = \,\,&(-1)^{l_1+l_2+L}\, \, C_{l_1}^{BB}C_{l_2}^{BB}\delta_{LL'}\delta_{MM'}\delta_{l_1l_4}\delta_{l_3l_2}\\ & + C_{l_1}^{BB}C_{l_2}^{BB}\delta_{LL'}\delta_{MM'}\delta_{l_2l_4}\delta_{l_3l_1},\\
\, \\
\bigg\langle {A}^{LM}_{l_1l_2|TE}{A}^{L'M'*}_{l_3l_4|TE} \bigg\rangle = \,\,&(-1)^{l_1+l_2+L}\, \,  C_{l_1}^{TE}C_{l_2}^{TE}\delta_{LL'}\delta_{MM'}\delta_{l_1l_4}\delta_{l_3l_2} \\ &+ C_{l_1}^{EE}C_{l_2}^{TT} \delta_{LL'}\delta_{MM'}\delta_{l_2l_4}\delta_{l_3l_1} ,\\
\, \\
\bigg\langle  {A}^{LM}_{l_1l_2|EB}{A}^{L'M'*}_{l_3l_4|EB}\bigg\rangle =\,\, &C_{l_1}^{EE}C_{l_2}^{BB}\delta_{LL'}\delta_{MM'}\delta_{l_2l_4}\delta_{l_3l_1},\\
\,\\
\bigg\langle {{A}^{LM}_{l_1l_2|TB}A}^{L'M'*}_{l_3l_4|TB} \bigg\rangle = \,\,&C_{l_1}^{BB}C_{l_2}^{TT} \delta_{LL'}\delta_{MM'}\delta_{l_2l_4}\delta_{l_3l_1} ,\\
\,\\
\bigg\langle {{A}^{LM}_{l_1l_2|TT}A}^{L'M'*}_{l_3l_4|EE} \bigg\rangle = \,\,&(-1)^{l_1+l_2+L} C_{l_1}^{TE}C_{l_2}^{TE} \delta_{LL'}\delta_{MM'}\delta_{l_2l_3}\delta_{l_4l_1} \\& +C_{l_1}^{TE}C_{l_2}^{TE} \delta_{LL'}\delta_{MM'}\delta_{l_2l_4}\delta_{l_3l_1},\\
\,\\
\bigg\langle {{A}^{LM*}_{l_1l_2|TT}A}^{L'M'}_{l_3l_4|TE} \bigg\rangle = \,\,&(-1)^{l_1+l_2+L} C_{l_1}^{TT}C_{l_2}^{TE} \delta_{LL'}\delta_{MM'}\delta_{l_2l_3}\delta_{l_4l_1} \\& +C_{l_1}^{TE}C_{l_2}^{TT} \delta_{LL'}\delta_{MM'}\delta_{l_2l_4}\delta_{l_3l_1},\\
\,\\
\bigg\langle {{A}^{LM*}_{l_1l_2|TE}A}^{L'M'}_{l_3l_4|EE} \bigg\rangle = \,\,&(-1)^{l_1+l_2+L} C_{l_1}^{TE}C_{l_2}^{EE} \delta_{LL'}\delta_{MM'}\delta_{l_2l_3}\delta_{l_4l_1} \\& +C_{l_1}^{EE}C_{l_2}^{TE} \delta_{LL'}\delta_{MM'}\delta_{l_2l_4}\delta_{l_3l_1}.
\end{split}
\end{align}
Variance of BipoSH coefficients for temperature was obtained by Hajian \cite{amir} and Joshi et al. \cite{nidhi}. For the BipoSH coefficients due to Doppler boost mentioned in  eq.\eqref{eq49}, \eqref{eq50}, \eqref{eq51}, the variance of BipoSH coefficients for polarization mentioned in eq.\eqref{eqes2} becomes,\\
\begin{align}\label{eqes3}
\begin{split}
&\bigg\langle {A}^{*10}_{ll+1|EE} {A}^{10}_{ll+1|EE}\bigg\rangle = C_{l}^{EE}C_{l+1}^{EE} ,\\
\,\\
&\bigg\langle {A}^{*10}_{ll+1|BB} {A}^{10}_{ll+1|BB}\bigg\rangle = C_{l}^{BB}C_{l+1}^{BB},\\
\, \\
&\bigg\langle {A}^{*10}_{ll+1|TE} {A}^{10}_{ll+1|TE}\bigg\rangle =  C_l^{EE}C_{l+1}^{TT},\\
\, \\
&\bigg\langle {A}^{10*}_{ll|EB} {A}^{10}_{ll|EB}\bigg\rangle =\,\, C_{l}^{EE}C_{l}^{BB},\\
\,\\
&\bigg\langle {A}^{10*}_{ll|TB} {A}^{10}_{ll|TB}\bigg\rangle = \,\,C_{l}^{BB}C_{l}^{TT},\\
\,\\
&\bigg\langle {A}^{*10}_{ll+1|TT} {A}^{10}_{ll+1|EE}\bigg\rangle = \,\, C_{l}^{TE}C_{l+1}^{TE},\\
\,\\
&\bigg\langle {A}^{*10}_{ll+1|TT} {A}^{10}_{ll+1|TE}\bigg\rangle = \,\, C_{l}^{TE}C_{l+1}^{TT},\\
\,\\
&\bigg\langle {A}^{*10}_{ll+1|TE} {A}^{10}_{ll+1|EE}\bigg\rangle = \,\, C_{l}^{EE}C_{l+1}^{TE}.
\end{split}
\end{align}
These are variance of BipoSH coefficients in absence of instrumental noise. In presence of instrumental noise, variance gets modified. The variance in eq.\eqref{eqes3} in the presence of instrumental noise can be obtained by replacing $C_l^{XX'}$ by $C_l^{XX'}+\mathcal{N}_{l}^{X}$, where $\mathcal{N}_{l}^{X}$ is the instrumental noise for $X= T, E, B$. Noise power spectrum $\mathcal{N}^{X}_l$  depends upon beam width and sensitivity of the instrument by \cite{marc,ftcmb,noise}, 
\begin{align}\label{eqes3a}
\begin{split}
&\mathcal{N}^{X}_l = \frac{\theta_{b}^2(\sigma_{pix}^2)_{X}}{n_{det}}{\rm e}^{l(l+1)\frac{\theta_{b}^2}{8ln2}}; \, \, \text{X= T, E,B}\\
&\mathcal{N}^{X}_l = \theta_{b}^2(\Sigma_{pix}^2)_{X}{\rm e}^{l(l+1)\frac{\theta_{b}^2}{8ln2}},
\end{split}
\end{align}
where, $\theta_{b}^2$ is FWHM of the Gaussian beam, $(\Sigma_{pix}^2)_{X}= \frac{(\sigma_{pix}^2)_{X}}{n_{det}}$ is the variance in $\theta_{b}^2$ pixel, $(\sigma_{pix}^2)_{X}$ is the variance in $\theta_{b}^2$ pixel per detector for temperature $(X=T)$ and polarization $(X=P)$, and $n_{det}$ are the number of detectors. For an equal integration time for two polarization states, pixel noise for temperature and polarization are related by \cite{marc},
\begin{align}\label{eqes3b}
(\sigma_{pix}^2)_{T}= \frac{(\sigma_{pix}^2)_{P}}{2}.
\end{align}
For $TE$ we have taken the instrumental noise as zero as mentioned by FUTURCMB \cite{ftcmb}. 
 

\section{Reconstruction noise for  Doppler boost from Planck and PRISM}\label{noi_es}
In the previous section we have obtained the expression for non-zero BipoSH coefficients, arising due to local motion.  In this section we derive the minimum variance reconstruction noise for polarization map for BipoSH coefficients. Instrumental noise for Planck is high, and is not suitable for making significant detection of $\boldsymbol\beta$ from polarization. However, noise level for future mission PRISM is smaller than Planck. This will make it possible to detect the magnitude and direction of $\beta$ from the polarization result. By measuring the non-zero BipoSH values from experiments like Planck and PRISM, we can infer the value of our local motion.  We have discussed the minimum variance estimator in more details in Sec. \ref{app3}.  Using eq.\eqref{eqnc8}, we estimate the value of reconstruction noise, $N^{XX'}_\beta$ for Planck and also for proposed future experiments like PRISM \cite{prismwhite} in the presence of instrumental noise, $\mathcal{N}^{X}_l$ as,

\begin{align}\label{eqes6}
\begin{split}
&N^{EE}_{1}= \left[\sum_{ll'}\frac{ S^{1}_{ll'|EE} S^{1}_{ll'|EE}}{2(C_{l}^{EE}+ \mathcal{N}^P_l)(C_{l'}^{EE} + \mathcal{N}^P_{l'})}\right]^{-1},\\
&N^{BB}_{1}= \left[\sum_{ll'}\frac{ S^{1}_{ll'|BB} S^{1}_{ll'|BB}}{2(C_{l}^{BB}+ \mathcal{N}^P_l)(C_{l'}^{BB} + \mathcal{N}^P_{l'})}\right]^{-1},\\
&N^{TE}_{ 1}= \left[\sum_{ll'}\frac{ S^{1}_{ll'|TE} S^{1}_{ll'|TE}}{C_{l}^{TE}C_{l'}^{TE} + (C_l^{EE}+\mathcal{N}^P_l)(C_{l'}^{TT}+ \mathcal{N}^T_{l'})}\right]^{-1},\\
&N^{EB}_{1}= \left[\sum_{ll}\frac{ S^{1}_{ll|EB} S^{1}_{ll|EB}}{(C^{EE}_{l}+ \mathcal{N}^P_l)(C^{BB}_{l} +\mathcal{N}^P_l)}\right]^{-1},\\
&N^{TB}_{1}= \left[\sum_{ll}\frac{ S^{1}_{ll|TB} S^{1}_{ll|TB}}{(C^{TT}_{l}+ \mathcal{N}^T_l)(C^{BB}_{l'}+ \mathcal{N}^P_{l'})}\right]^{-1},\\
&N^{TT}_{1}= \bigg[\sum_{ll'}\frac{ S^{1}_{ll'|TT} S^{1}_{ll'|TT}}{2(C^{TT}_{l}+ \mathcal{N}^T_{l})(C^{TT}_{l'}+ \mathcal{N}^T_{l'})}\bigg]^{-1},
\end{split}
\end{align}

where shape factor $ S^{1}_{ll'}$ are,
\begin{align}\label{eqes5}
\begin{split}
& S^{1}_{ll+1|EE}= 2\pi \, \frac{D^{EE}_l }{\beta_{10} l(l+1)}\frac{\Pi_{l\,l+1}}{\sqrt{4\pi}\Pi_{1}}C^{1 0}_{l+1\,- 2\, l\, 2}\,,\\
& S^{1}_{ll+1|BB}= 2\pi \, \frac{D^{BB}_l }{\beta_{10} l(l+1)}\frac{\Pi_{l\,l+1}}{\sqrt{4\pi}\Pi_{1}}C^{1 0}_{l+1\,- 2\, l\, 2}\,,\\
& S^{1}_{ll+1|TE}= 2\pi \, \frac{D^{TE}_l }{\beta_{10} l(l+1)}\frac{\Pi_{l\,l+1}}{\sqrt{4\pi}\Pi_{1}}C^{1 0}_{l+1\,- 2\, l\, 2}\,,\\
& S^{1}_{ll|EB}= 2\pi \, \frac{D^{EB}_l }{\beta_{10} l(l+1)}\frac{\Pi_{l}^2}{\sqrt{4\pi}\Pi_{1}}C^{1 0}_{l\,- 2\, l\, 2}\,,\\
& S^{1}_{ll|TB}= 2\pi \, \frac{D^{TB}_l }{\beta_{10} l(l+1)}\frac{\Pi_{l}^2}{\sqrt{4\pi}\Pi_{1}}C^{1 0}_{l\,- 2\, l\, 2}\,,\\
&S^{1}_{ll+1|TT}= 2\pi \, \frac{D^{TT}_l }{\beta_{10} l(l+1)}\frac{\Pi_{l\,l+1}}{\sqrt{4\pi}\Pi_{1}}C^{1 0}_{l+1\,0\, l\, 0}\,.
\end{split}
\end{align}
The value of the reconstruction noise sets the maximum possible significance of the measurement of $\beta$ from the corresponding observable. For the theoretical best-fit value of $C^{XX'}_{l}$, with no instrumental noise, leads to minimum  reconstruction noise. This is the theoretical noise  due to the cosmic variance, which is inevitable. But for any experiment, the variance is enhanced by instrumental sensitivity and finite angular resolution. Presence of instrumental noise leads to increase in the reconstruction noise, $N^{XX'}_1$. As a result of which  measuring the value of $\beta$ becomes difficult. \\

\begin{figure}[H]
\centering
\includegraphics[width=6.0in,keepaspectratio=true]{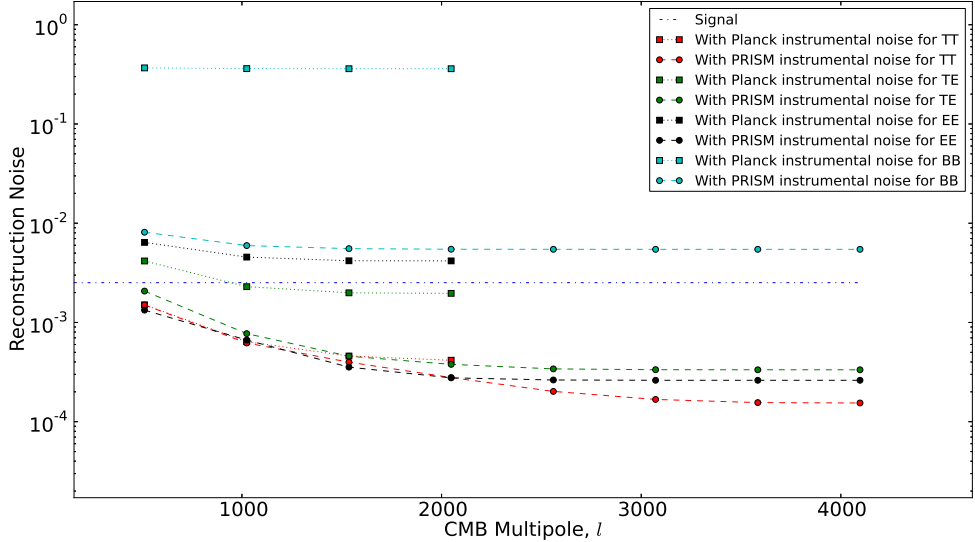}
\caption{Cumulative binned reconstruction noise for $TT$, $EE$, $TE$ and $BB$ with $\Delta l= 512$. We have taken instrumental noise for Planck (dash-square), and PRISM(dash-circle). Doppler boost signal, $\beta_{10}= 2.52 \times 10^{-3}$ (dot-dash). Here we have used the best-fit $\Lambda$CDM $C_l^{EE}, C_l^{BB}$, $C_l^{TE}$ and $C_l^{TT}$ generated from CAMB\cite{camb} with tensor to scalar ratio $(r)= 0.10$ and value of $b_\nu \approx 3$ for $\nu = 217$ GHz and $220$ GHz for Planck\, and PRISM respectively.}\label{fig:f}
\end{figure}

\begin{figure}[H]
\centering
\includegraphics[width=6.0in,keepaspectratio=true]{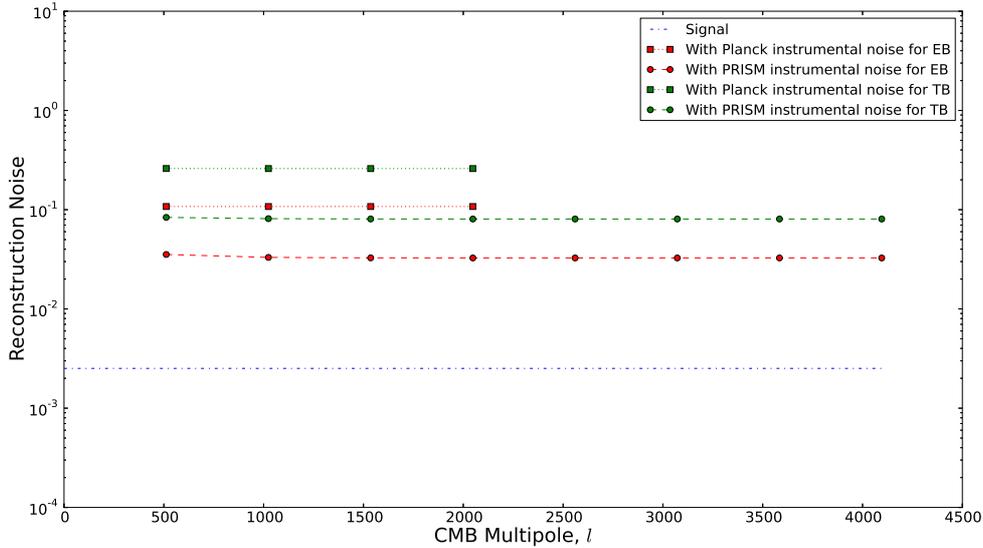}
\caption{Cumulative binned reconstruction noise for $EB$ and $TB$ with $\Delta l= 512$. We have taken instrumental noise for Planck (dash-square), and PRISM(dash-circle). Doppler boost signal, $\beta_{10}= 2.52 \times 10^{-3}$ (dot-dash).     Here we have used the best-fit $\Lambda$CDM $C_l^{EE}, C_l^{BB}$, $C_l^{TE}$ and $C_l^{TT}$ generated from CAMB\cite{camb} with tensor to scalar ratio $(r)= 0.10$ and value of $b_\nu \approx 3$ for $\nu = 217$ GHz and $220$ GHz, for Planck\, and PRISM respectively.}\label{fig:g}
\end{figure}

For Planck, we made the estimation for the frequency channel, $\nu= 217$ GHz, with the following detector characteristics \cite{planckblue}, 
\begin{enumerate} 
\item
$\theta_{FWHM}= 5 \,\,\text{arcminute}$. 
\item
$(\Sigma_{pix})_{P} = 9.8 \mu K/K$.
\item
$(\Sigma_{pix})_{T} = 4.8 \mu K/K$. 
\end{enumerate}
Similar to Planck, we also estimate the reconstruction noise, $N_{\beta}$ for future experiment, PRISM.  From the PRISM white paper \cite{prismwhite}, we  obtain the instrumental noise  for $\nu=220$ GHz with instrumental characteristics,
\begin{enumerate} 
\item
$\theta_{FWHM}= 2.3' \,\,\text{arcminute}$. 
\item
$(\sigma_{pix})_{P} = 71.9 \mu K/K$.
\item
$(\sigma_{pix})_{T} =  50.9 \mu K/K$.
\item
$n_{det}= 350$.
\end{enumerate}
Since PRISM proposes to achieve much higher angular resolution than Planck, and also much lower instrumental noise, we expect the reconstruction noise for PRISM to be smaller than that obtained for Planck. The comparison between  reconstruction  noise eq.\eqref{eqes6} for Planck and PRISM 
are given in fig.~\ref{fig:f} for $TT$, $EE$, $BB$ and $TE$, and in fig.\ref{fig:g} for $EB$ and $TB$. We used the best-fit parameters from Planck \cite{Planck3} for the value of $C_{l}^{XX'} (X= T, E, B)$ to evaluate eq.\eqref{eqes1}. We plotted $(N_1)^{1/2}$ in fig.~\ref{fig:f},\,\ref{fig:g}  along with $\beta_{10}= 2.52 \times 10^{-3}$.
In principle, $\beta$ is measurable  in all the BipoSH spectra, except for $EB$ and $TB$. We summarise the possibility of detection of $\beta$ for Planck and PRISM, from the calculated reconstruction noise, plotted in fig.\ref{fig:f} and \ref{fig:g}. 
\begin{enumerate} 
\item
Reconstruction noise for $EE$ and $BB$ with Planck instrumental noise and angular resolution with tensor to scalar ratio $(r)= 0.11$ is much above the signal $(\beta_{10}= 2.53 \times 10^{-3})$, as shown by curve (dash-square)  in fig.\ref{fig:f}. This implies we cannot make statistically significant detection of $\beta$ in $EE$ and $BB$  from Planck.\\
\item
The detection of $\beta$ is consistent with our expectation, in $TT$. A $6 \sigma$ detection of $\beta$ is possible at $l_{max} =2048$ and $4\sigma$ detection is possible at $l_{max}= 1024$ in frequency channel, $\nu= 217$ GHz with corresponding $b_{\nu}= 3$.
 A $3 \sigma$  detection has been already reported by Planck \cite{Planck1} at $l_{max}=2048$. Kosowsky \& Kahniashvili \cite{kosowsky} and Amendola et al.\cite{amendola} have  also mentioned a Signal to Noise Ratio (SNR)  of  $6$ and $4$ respectively for $\beta$ from Planck for frequency channel $\nu = 143$ GHz and have taken the value of $b_{\nu} =1$. Using BipoSH formalism we have also recovered a similar SNR for $\nu= 143$ GHz. 
\item
A measurement of $\beta$  from $TT$ BipoSH coefficients with a significance of $15\sigma$ at $l_{max}= 4096$ and $4.1\sigma$ detection is possible at $l_{max}= 1024$  with PRISM in the frequency channel $\nu= 220$ GHz. However, presence of other secondary effects and foreground contaminations can be expected to reduce the significance of the detection for  $l_{max}$ values beyond $\sim 3000$.\\
\item 
Planck can lead to weak detection of $\beta$ from $TE$ spectra with a $1.3\sigma$ significance at $l_{max}= 1500$  for $\nu = 217$ GHz. Amendola et al. \cite{amendola} have shown a possibility of measuring $\beta$ from $\nu= 143$ GHz channel with $b_{\nu}= 1$ with SNR of $2$. In BipoSH formalism, we recover a $1.95\sigma$ detection of $\beta$ for $\nu= 143$ GHz. But using PRISM we can make a measurement of $\beta$   from both $EE$ and $TE$ with a significance of $9.4\sigma$ and $7.4\sigma$ respectively at $l_{max}= 4096$ and $3.8\sigma$ and $3.3\sigma$ respectively at $l_{max}= 1024$ in the frequency channel $\nu= 220$ GHz ignoring the contamination from other foregrounds.\\
\item
In $BB$ BipoSH coefficients, measurement of $\beta$ is not possible by PRISM. This is due to low value of $C_l^{BB}$ signal relative to the instrumental noise. The theoretical reconstruction noise for $BB$ is $0.76 \times 10^{-3}$ which clearly indicates the possibility of detecting $\beta$ from experiments with better resolution and sensitivity than PRISM.\\
\item
Measurement of $\beta$ from $EB$ and $TB$ BipoSH coefficients are not possible even in principle. The reconstruction noise for PRISM are much above the value of $\beta$ fig.\ref{fig:g}. Due to negligible contribution to $B$ modes from $E$ modes at high $l$, the reconstruction noise gets saturated at lower $l$. The reconstruction noise can be improved by combining the reconstruction noise of all the BipoSH spectra.
\end{enumerate}
Our result shows us that we can estimate $\beta$ with high significance from polarization maps from PRISM, which is not possible from the experiments like Planck.


\section{Conclusion}
The local motion of the observer as measured from CMB dipole \cite{WMAP} (with $\beta \equiv v/c= 1.23 \times 10^{-3}$) causes an unavoidable violation of Statistical Isotropy effect on the CMB temperature and polarization field. Recently, Planck \cite{Planck1} has shown that Doppler boost of temperature fluctuation leads to mixing of power between different CMB multipoles $l$, which results in non-zero value of the off-diagonal (BipoSH) terms in the SH space covariance matrix. The Doppler boost of temperature fluctuation results in violation of Statistical Isotropy (SI) in the observer's frame. This has been recently measured in the temperature map in the Planck results.  Even more, recently Jeong et al. \cite{jeong} showed that the effect of aberration from boost on temperature angular power spectrum on partial sky measurements could be responsible for the differences in Planck and SPT measurements at high multipoles.\\
\\
In this paper we have discussed the effect of Doppler boost on the CMB polarization field. Boost affects  the CMB polarization field by modulation of the Stokes parameters and aberration in the direction of incoming photons. Because of these two effects, both diagonal and off-diagonal terms in SH space covariance matrix of polarization field get affected. The leading order effect to angular power spectrum is $O(\beta^2)$ as mentioned in eq.\eqref{eq44}, eq.\eqref{eq44a} and eq.\eqref{eq43b}. This means, angular power spectra are mildly affected by boost at low $l$. But the off-diagonal terms of the covariance matrix in SH space, have  effect linear in  $\beta$. This  leads to the non-zero potentially measurable BipoSH spectra ( $A^{10}_{ll_+1|XX'}$) as mentioned in eq.\eqref{eq49}, eq.\eqref{eq50} and eq.\eqref{eq51} and plotted in fig.\ref{fig:e}. This implies, a SI CMB polarization field would appear Non-SI (nSI) in our frame. The measurement of these non-zero BipoSH coefficients  lead to an estimation of ${\beta}$. Statistically significant measurement of  $\beta$  depends upon the reconstruction noise of any future experiments like Planck and PRISM. We have estimated the reconstruction noise for  Planck and PRISM by using  minimum variance estimator method. Reconstruction noise depends upon the variance of the BipoSH coefficients and instrumental noise. The variance of BipoSH coefficients for polarization are calculated, and using the instrumental noise of Planck and PRISM, we have estimated the reconstruction noise for $TT$, $EE$, $TE$ and $BB$. $\beta$ is detectable with a high significance in $TT$, poorly in $TE$ for Planck and in $TT$, $EE$, $TE$ for PRISM. But $\beta$ is not measurable in $BB$, $EB$ and $TB$ BipoSH coefficients by Planck and PRISM. The theoretical reconstruction noise for $BB$ is smaller than $\beta$. But due to higher value of instrumental noise relative to the signal, detection of $\beta$ from $BB$ spectra is not possible. Whereas, $EB$ and $TB$ spectra can never be used for measuring the value of $\beta$ using minimum variance estimator because of high theoretical reconstruction noise. The results here can be used to derive similar implication for comparing CMB polarization angular power spectra measured by CMB experiments covering small patches of the sky as shown for temperature anisotropy in \cite{jeong}. The computed BipoSH spectra from the Doppler boost need to be accounted in all future SI violation searches from the upcoming temperature and polarization maps such as Planck and PRISM. \\

\textbf{Acknowledgement:}\\
SM acknowledges Council for Science and Industrial Research (CSIR), India, for the financial support as Senior Research Fellow. TS acknowledges support from Swarnajayanti fellowship, DST, India. TS thank Anthony Challinor for pointing to an important reference. We also thank Aditya Rotti and Santanu Das for many useful discussions. 

\begin{appendix}
\section{Leading second order corrections to the angular power spectra}\label{app2}
 Here we provide details of the calculation of the leading order effect at $O(\beta^2)$ on the angular power spectra. By taking the product of $_+ \tilde{X}_{l_1m_1}$  defined in eq.\eqref{eq25} with $_+ \tilde{X}^*_{l_2m_2}$ and with $_- \tilde{X}^*_{l_2m_2}$, and retaining only $O(\beta^2)$ terms, we get
 
\begin{align}\label{eq43x}
\begin{split}
\tilde{C}^{EE}_{l_1} =&\,\, C^{EE}_{l_1}  + \frac{1}{2}\beta^2 \Bigg[ \sum_{l_2m_2}\sum_{M} {}_{+2}I^{m_1m_2M}_{l_1l_21} \left[\left(C^{EE}_{l_2} + C^{BB}_{l_2}\right) {}_{+2}I^{m_1m_2M} _{l_1l_21} + \left(C^{EE}_{l_2} - C^{BB}_{l_2}\right) {}_{-2}I^{m_1m_2}_{l_1l_21} \right] \\& + \sum_{M}  \left[ \left(C^{EE}_{l_1} + C^{BB}_{l_1}\right) {}_{+2}J^{m_1m_1MM}_{l_1l_1 1 1} + \frac{1}{2} \left(C^{EE}_{l_1} - C^{BB}_{l_1}\right)\left[_{+2}J^{m_1m_1MM}_{l_1l_1 1 1}+ {}_{-2}J^{m_1m_1MM}_{l_1l_1 1 1}\right] \right] \\& + \left(\ b_\nu \right)^2 \sum_{M}\sum_{l_2m_2}  {}_{+2}H^{m_1m_2M}_{l_1l_21} \left[ \left(C^{EE}_{l_2} + C^{BB}_{l_2}\right) {}_{+2}H^{m_1m_2M}_{l_1l_21} + \left(C^{EE}_{l_2} - C^{BB}_{l_2}\right) {}_{-2}H^{m_1m_2M}_{l_1l_21}\right]\\& -  2b_\nu \sum_{M}\sum_{l_2m_2}  {}_{+2}H^{m_1m_2M}_{l_1l_21} \left[\left(C^{EE}_{l_2} + C^{BB}_{l_2}\right) {}_{+2}I^{m_1m_2M}_{l_1l_21} + \left(C^{EE}_{l_2} - C^{BB}_{l_2}\right) {}_{-2}I^{m_1m_2M}_{l_1l_21}\right]  \\& - b_\nu \sum_{M}  \left[2 \left(C^{EE}_{l_1} + C^{BB}_{l_1}\right) {}_{+2}G^{m_1m_1MM}_{l_1l_1 1 1} +  \left(C^{EE}_{l_1} - C^{BB}_{l_1}\right)\left[_{+2}G^{m_1m_1MM}_{l_1l_1 1 1}+ {}_{-2}G^{m_1m_1MM}_{l_1l_1 1 1}\right] \right]\Bigg],
\end{split}
\end{align}
similarly we can obtain the expression for $C_l^{BB}$ as,
\begin{align}\label{eq44y}
\begin{split}
\tilde{C}^{BB}_{l_1} =& \,\,C^{BB}_{l_1}  + \frac{1}{2}\beta^2 \Bigg[\sum_{l_2m_2}\sum_{M} {}_{+2}I^{m_1m_2M}_{l_1l_21} \left[\left(C^{EE}_{l_2} + C^{BB}_{l_2}\right) {}_{+2}I^{m_1m_2M} {}_{l_1l_21} - \left(C^{EE}_{l_2} - C^{BB}_{l_2}\right) {}_{-2}I^{m_1m_2M}_{l_1l_21} \right] \\& + \sum_{M} \left[\left(C^{EE}_{l_1} + C^{BB}_{l_1}\right) {}_{+2}J^{m_1m_1MM}_{l_1l_1 1 1} - \frac{1}{2} \left(C^{EE}_{l_1} - C^{BB}_{l_1}\right)\left[_{+2}J^{m_1m_1MM}_{l_1l_1 1 1}+ {}_{-2}J^{m_1m_1MM}_{l_1l_1 1 1}\right] \right] \\& + \left( b_\nu \right)^2 \sum_{M}\sum_{l_2m_2}  {}_{+2}H^{m_1m_2M}_{l_1l_21} \left[\left(C^{EE}_{l_2} + C^{BB}_{l_2}\right) {}_{+2}H^{m_1m_2M}_{l_1l_21} - \left(C^{EE}_{l_2} - C^{BB}_{l_2}\right) {}_{-2}H^{m_1m_2M}_{l_1l_21}\right] \\& -  2b_\nu \sum_{M}\sum_{l_2m_2}  {}_{+2}H^{m_1m_2M}_{l_1l_21} \left[\left(C^{EE}_{l_2} + C^{BB}_{l_2}\right) {}_{+2}I^{m_1m_2M}_{l_1l_21} - \left(C^{EE}_{l_2} - C^{BB}_{l_2}\right) {}_{-2}I^{m_1m_2M}_{l_1l_21}\right] \\& - b_\nu \sum_{M} \left[2\left(C^{EE}_{l_1} + C^{BB}_{l_1}\right) {}_{+2}G^{m_1m_1MM}_{l_1l_1 1 1} -  \left(C^{EE}_{l_1} - C^{BB}_{l_1}\right)\left[_{+2}G^{m_1m_1MM}_{l_1l_1 1 1}+ {}_{-2}G^{m_1m_1MM}_{l_1l_1 1 1}\right] \right]\Bigg].
\end{split}
\end{align}

The $TE$ correlation can be obtained as, 

\begin{align}\label{eq44z}
\begin{split}
\tilde{C}^{TE}_{l_1} =&\,\, C^{TE}_{l_1} +\beta^2\Bigg[ \sum_{l_2m_2}\sum_{M}C^{TE}_{l_2} {}_{+2}I^{m_1m_2M}_{l_1l_21} {}_{0}I^{m_1m_2M}_{l_1l_21}  + \sum_{M}\frac{1}{2}C^{TE}_{l_1} {}_{+2}J^{m_1m_1MM}_{l_1l_1 1 1} \\& + \sum_{M}\frac{1}{2}C^{TE}_{l_1} {}_{0}J^{m_1m_1MM}_{l_1l_1 1 1} + \left( b_\nu \right)^2 \sum_{M}\sum_{l_2m_2}C^{TE}_{l_2} {}_{+2}H^{m_1m_2M}_{l_1l_21} {}_{0}H^{m_1m_2M}_{l_1l_21} \\& - b_\nu \bigg( \sum_{M}\sum_{l_2m_2}C^{TE}_{l_2} {}_{+2}H^{m_1m_2M}_{l_1l_21} {}_{0}I^{m_1m_2M}_{l_1l_21} +  \sum_{M}\sum_{l_2m_2}C^{TE}_{l_2} {}_{+2}
I^{m_1m_2M}_{l_1l_21}{}{}_{0}H^{m_1m_2M}_{l_1l_21}\\&  \sum_{M}C^{TE}_{l_1} {}_{+2}G^{m_1Mm_1M}_{l_1 1 l_1 1} +   \sum_{M}C^{TE}_{l_1}  {}_{0}G^{m_1Mm_1M}_{l_1 1 l_1 1}\bigg) \Bigg].
\end{split}
\end{align}

After summing over $m_2, M$ by using the properties of  Clebsch-Gordan coefficients in eq.\eqref{eq43x}, \eqref{eq44y}, \eqref{eq44z}, we can get,

\begin{align}\label{eq28c}
\begin{split}
\sum_{m_2M} {}_{+2}H_{l_1l_21}^{m_1m_2M} {_{+2}H_{l_1l_21}^{m_1m_2M}}=& \,\, \frac{\Pi_{l_2}^2}{4\pi} (C^{10}_{l_2\, 2 \,l_1 \,-2})^2,\\
\sum_{m_2M} {}_{-2}H_{l_1l_21}^{m_1m_2M} {_{+2}H_{l_1l_21}^{m_1m_2M}}=& \,\,(-1)^{l_1+l_2+ 1} \frac{\Pi_{l_2}^2}{4\pi} (C^{10}_{l_2\, 2\, l_1\, -2})^2,\\
\sum_{m_2M} {}_{+2}I_{l_1l_21}^{m_1m_2M} {_{+2}I_{l_1l_21}^{m_1m_2M}}=& \,\,\frac{\Pi_{l_2}^2}{16\pi} \bigg[l_2(l_2+1)+2 - l_1(l_1+1)\bigg]^2(C^{10}_{l_2\, 2 \,l_1\, -2})^2,\\
\sum_{m_2M} {}_{-2}I_{l_1l_21}^{m_1m_2M} {_{+2}I_{l_1l_21}^{m_1m_2M}}=&\,\, (-1)^{l_1+l_2+ 1} \frac{\Pi_{l_2}^2}{16\pi} \bigg[l_2(l_2+1)+2 - l_1(l_1+1)\bigg]^2(C^{10}_{l_2\, 2\, l_1\, -2})^2.
\end{split}
\end{align}
 Also, the integral in $_{\pm 2}G^{m_1m_1m_3M}_{l_1l_1 1 1}$ can be obtained by using the properties of the spherical harmonics from \cite{varsha},
\begin{align}\label{eq45}
\begin{split}
&\sum_{M} {}_{\pm 2}G^{m_1m_2m_3M}_{l_1l_2 l_3 L}= \sum_{J} \frac{\Pi_{J}^2 \bigg[l_3(l_3+1) +l_2(l_2+1) - J(J+1)\bigg]}{8\pi} C^{l_3 0}_{l_2\, \pm 2 \,J \,\mp 2}C^{L 0}_{l_1\,\pm 2 \,J \,\mp 2} \delta_{l_1 l_2} \delta_{m_1 m_2},\\
&\sum_{M}  {}_{0}G^{m_1m_2m_3M}_{l_1l_2 l_3 L}= \sum_{J} \frac{\Pi_{J}^2\bigg[l_3(l_3+1) +l_2(l_2+1) - J(J+1)\bigg]}{8\pi} C^{l_3 0}_{l_2 0 J 0}C^{L 0}_{l_1 0 J 0} \, \delta_{l_1 l_2} \, \delta_{m_1 m_2},
\end{split}
\end{align}
and the value of the $\sum_{M} {}_{+2}J^{m_1m_1MM}_{l_1l_1 1 1}$ can be obtained from the \cite{wayne} as, 
\begin{align}\label{eq46}
\begin{split}
\sum_{M} {}_{0}J^{mmMM}_{ll L L}=& \int d \hat n Y^*_{lm} \bigtriangledown_i Y_{LM}(\hat n) \bigtriangledown_j Y^*_{L M}(\hat n)  \bigtriangledown^i \bigtriangledown ^j Y_{lm}(\hat n),\\
=& -\frac{\Pi_{L}^{2}}{8\pi}lL(l+1)(L+1),
\end{split}
\end{align}

\begin{align}\label{eq46a}
\begin{split}
\sum_{M} {}_{\pm 2}J^{mmMM}_{ll L L}=& \int d \hat n\, {}_{\pm2}Y^*_{l m} \bigtriangledown_i Y_{LM}(\hat n) \bigtriangledown_j Y^*_{L M}(\hat n)  \bigtriangledown^i \bigtriangledown ^j {}_{\pm2}Y_{lm}(\hat n),\\
=& -\frac{\Pi_{L}^{2}}{8\pi}L(L+1)\bigg[l(l+1)-4\bigg].
\end{split}
\end{align}

The final results obtained using these calculations are given in eqs.\eqref{eq44}, \eqref{eq44a}, \eqref{eq43b} in Sec.~\ref{po2}. 

\end{appendix}

\end{document}